\documentclass[prd,superscriptaddress,nofootinbib]{revtex4}
\usepackage{amssymb,amsmath,graphicx,amscd,epsfig}
\usepackage[colorlinks=true, pdfstartview=FitV, linkcolor=blue, citecolor=red, urlcolor=magenta, breaklinks=true]{hyperref}

\usepackage[T1,T5]{fontenc}
\usepackage{graphicx}
\usepackage[all]{xy}
\usepackage{amsmath}
\usepackage{amssymb}
\usepackage{color}
\usepackage{subeqnarray}
\usepackage{subfigure}
\usepackage{epstopdf}

\newcommand{\bb}{\bibitem}
\newcommand{\bes}{\begin{subequations}}
\newcommand{\ees}{\end{subequations}}
\def\ben{\begin{eqnarray}}
\def\een{\end{eqnarray}}
\newcommand{\bens}{\begin{subeqnarray}}
\newcommand{\eens}{\end{subeqnarray}}

\def\be{\begin{equation}}
\def\ee{\end{equation}}
\def\e{\text{e}}
\def\o{\text{o}}
\def\tanh{\text{tanh}}
\def\sech{\text{sech}}

\def\cos{\text{cos}}

\begin{document}

\title{Scalar field models driven by Dirac-Born-Infeld dynamics and their relatives}

\author{Elisama E. M. Lima} \email{elisama.lima@ifba.edu.br}
\affiliation{Federal Institute of Education, Science and Technology of Bahia, Barreiras, Bahia, 47808-006, Brazil}

\author{Francisco A. Brito}\email{fabrito@df.ufcg.edu.br}
\affiliation{Departamento de F\'{\i}sica, Universidade Federal de Campina Grande
Caixa Postal 10071, 58429-900 Campina Grande, Para\'iba, Brazil}
\affiliation{Departamento de F\'isica, Universidade Federal da Para\'iba, 
Caixa Postal 5008, 58051-970 Jo\~ao Pessoa, Para\'iba, Brazil}

	




\begin{abstract}
In this paper, we investigate novel kinklike structures in a scalar field theory driven by Dirac-Born-Infeld (DBI) dynamics. Analytical features are reached through a first-order formalism and a deformation procedure. The analysis ensures the linear stability of the obtained solutions, and the deformation method permits to detect new topological solutions given some systems of known solutions. The proposed models vary according to the parameters of the theory. However, in a certain parameter regime, their defect profiles are precisely obtained by standard theories. These are the models relatives. Besides that, we investigate the $\beta-$Starobinsky potential in the perspective of topological defects; and we have shown that it can support kinklike solutions, for both canonical and non-canonical kinetics. As a result, we propose two new kinds of generalizations on the $\beta-$Starobinsky model, by considering the DBI approach. Finally, we explore the main characteristics of such structures in these new scenarios. 
\end{abstract}

\maketitle

\pretolerance10000

\section{Introduction} \label{intro}

The Dirac-Born-Infeld scalar theory was inspired by the non-linear electromagnetism \cite{Born:1934gh,Dirac:1962iy}, which coincides with the standard theory for small field excitations.  Initially, the theory was proposed to eliminate divergences in the classical electrodynamics 
and after, it also proved useful to investigate topological structures in non-ordinary field contexts \cite{Brown,Sarangi,Babichevkvortices,Andrews,Garcia,Bazeia:2017mnc}. For example, it can support instanton solutions \cite{Brown}, global string solutions  \cite{Sarangi}, vortex solutions \cite{Babichevkvortices}, besides other kinds of topological defects \cite{Andrews,Garcia,Bazeia:2017mnc,tops}.
 
The present work considers novel scalar field models with DBI-like kinetic term, and we analyze how that generalization changes the topological solutions. In this situation, the equations of motion include non-linear terms that can be diminished by a first-order framework \cite{FOFGD}, and the linear stability of the solutions is ensured. For some cases, the non-linearities provoked by the modified dynamics mimic an inner structure, featured by a plateau on the stability potential \cite{brane}. 

The novel analytical models are discovered through the deformation procedure, which consists of proposing a deforming function connecting two potentials, one of the known solutions and another unknown which one desires to find \cite{Bazeia:2002xg,DMGD}. The Dirac-Born-Infeld kink like solutions, in a certain parameter regime, encodes the same field profile of ordinary scalar theories; in such a way, the energy density and the stability potential of their canonical counterparts are precisely recovered \cite{Bazeia:2017mnc}. That is the reason why we are calling these models relatives.

The models and their relatives can be useful to provide both the topological and cosmological inflaton solutions. While in the first case we look for static solutions with spatial dependence, in the second case we look for homogeneous solutions, i.e., time dependent only solutions. From the inflationary cosmology point of view, we need scalar dynamics that can roll slowly enough to get the precise amount of inflation to solve the horizon, flatness and monopole problem.  One of the main problems in string theory or supergravity is to find such potentials able to develop this precise behavior. One has found several potentials in these contexts that are are easily ruled out by the cosmological observational data. Thus several other approaches keeping some characteristics around fundamental theories as brane inflation \cite{dvali-tye}, flux compactifications \cite{KKLT}, to quote a few have been addressed in the literature. On the other hand, the DBI dynamics for being directly related to brane dynamics seems to be a good place to look for potentials that can overcome such difficulties. Same can be said to their deformed counterparts since the deformation can make an initial unsuitable potential to a final completely acceptable potential to describe inflationary cosmology. In this sense we consider one of the most celebrated potentials to describe dynamics of inflation, the Starobinsky potential, to address its deformation in the DBI dynamics. As we shall show, their relatives may find applications in both cosmology and (non-)topological solutions.

The paper is organized as follows. In Sec.~\ref{sec-1} we introduce generalities about the first-order formalism and deformation mechanism. Then, we focus on the standard dynamics in Sec.~\ref{sec-2} and DBI dynamics in Sec.~\ref{sec-3} in the context of deformed theories. The final considerations are presented in Sec.~\ref{sec-com}.

\section{General description} \label{sec-1}

In this section, we shall describe general properties of the formalism used in this work, to characterize a real scalar field in a non-standard theory, as shown in Refs.~\cite{FOFGD}. The generalized action in the $(1+1)-$dimensional space-time is 
\be
\label{lagran0}
S=\int d^2x {\cal L}\left(\phi, X \right),
\ee
where $X=\frac12\partial_\mu\phi\partial^\mu\phi$. For the specific case where the Lagrangian density is of the type
\be
\label{lagran}
{\cal L}\left(\phi,X\right)=F(X)-V(\phi);
\ee
the equation of motion reads
\be
\partial_\mu(F_{X}\partial^\mu\phi)=-V_{\phi};
\label{eom0}
\ee
and the energy-momentum tensor is
\be
\label{tensor0}
T_{\mu\nu}=F_{X}\partial_\mu\phi\partial_\nu\phi-g_{\mu\nu}{\cal{L}}.
\ee

Since we are interested in topological solutions, one considers static field configurations. Then the equation of motion reduces to the form
\be
(2XF_{XX}+F_{X})\phi''=V_\phi.
\label{eom1}
\ee
The energy-momentum components $T_{00}$ and $T_{ii}$ display the energy density and stress of the solution, respectively:
\begin{eqnarray}
\rho(x)&=&-{\cal{L}}\\
\tau(x)&=&F_{X}\phi'^2+{\cal{L}}.
\label{stress}
\end{eqnarray}
Assuming the stressless condition required by stability reasons, $F_{X}\phi'^2+{\cal{L}}=0$. The energy density will be written in the form 
\be
\rho(x)=-{\cal{L}}=F_{X}\phi'^2.
\label{rho}
\ee

It is useful to introduce the first-order framework so that the equation of motion is satisfied. For that reason, we define a function $W=W(\phi)$ such as
\be
F_{X}\phi'=W_{\phi}.
\label{w}
\ee
The second-order equation, $(F_{X}\phi')'=V_{\phi}$, requires that the function $W$ also obeys 
\be
W_{{\phi}{\phi}}\phi'=V_{\phi}.
\label{here}
\ee
It means that $F_X V_\phi=W_\phi W_{\phi\phi}$, which is verified for the potential under the stressless condition, $V(\phi)=F(\phi')+W_{\phi}(\phi)\phi'$. 

The energy density assumes the form
\be
\rho(x)=W_\phi \phi'=\frac{d W}{dx},
\label{rhoFx}
\ee
and the total energy can be given by the function $W$ at the asymptotic limits 
\begin{eqnarray}
E=|\Delta W|&=&|W(\phi(x\rightarrow \infty))-W(\phi(x\rightarrow -\infty))|. \nonumber \\
\end{eqnarray}
The standard situation is recovered assuming $F(X)=X$ \cite{bogomol}.

\subsection{Linear stability}

The analysis of the linear stability is made assuming small time-dependent perturbations around the static solution: $\phi(x,t)=\phi_s(x)+\eta_n(x)\cos(\omega_n t)$. The linear perturbations lead to
\ben
X&=&-\frac12\phi_s'^2-\phi_s'\eta'_n\cos(\omega_n t) \\
&=&X_s+\overline{X}
\een
with $\overline{X}=-\phi_s'\eta'_n\cos(\omega_n t)$. The equation of motion \eqref{eom0} becomes
\be
\label{eqpert}
-\left[\left(F_{X_s}+2X_s F_{X_s X_s}\right)\eta'_n\right]'= \left( F_{X_s}\omega_n^2-V_{\phi_s\phi_s}\right)\eta_n,
\ee
which is a  generalized Sturm-Liouville (SL) equation. 

The Liouville normal form can be obtained, given the conditions: $A^2F_{X_s}>0$ and $F_{X_s}>0$, where 
\be
A^2=\frac{2X_s F_{X_s X_s}+F_{X_s}}{F_{X_s}}.
\label{A}
\ee
Besides, for non-singular SL problems, $A^2F_{X_s}$ and $F_{X_s}$ must be continuously differentiable, and $V_{\phi_s\phi_s}$  is required to be continuous. 

Applying the Liouville transformation
\begin{eqnarray}
\label{transformation}
dz=\frac{1}{A(x)}dx \qquad \mbox{and}\qquad
\varepsilon_n=\sqrt{F_{X_s}A}\,\eta_n.
\end{eqnarray}
The SL equation transforms into an eigenvalue problem similar to the one-dimensional time-independent Schr\"odinger equation, 
\be
\left(-\frac{d^2}{dz^2}+U(z)\right)\varepsilon_n(z)=\omega^2\varepsilon_n (z),
\label{schrodingerlikequation}
\ee
where the stability potential is given by
\be
U(z)=\left. \frac{\sqrt{F_{X_s}A}}{F_{X_s}}\frac{d}{dx}\left(A \frac{d \left(\sqrt{F_{X_s}A}\right)}{dx}  \right)
+ \frac{V_{\phi_s\phi_s}}{F_{X_s}} \,  \right|_{x=x(z)}.
\label{stabilitypotential}
\ee
The zero mode related to $\omega_0=0$ gives
\be
U(z)=\frac{1}{\varepsilon_0(z)}\frac{d^2}{dz^2}\varepsilon_0(z).
\ee
Comparing the SL equation for $\omega_0=0$ that is  $-(A^2F_{X_s}\eta'_0)'+V_{\phi_s\phi_s}\eta_0=0$ and the derivative of the static motion equation \eqref{eom1} that is $-(A^2F_{X_s}\phi'')'+V_{\phi_s\phi_s}\phi'=0$;  we note that the zero mode is $\eta_0=\phi'$. Therefore, linear stability potential can also be written as
\be
U(z)=\left. \frac{\sqrt{F_{X_s} A}}{\phi' F_{X_s}}\frac{d}{dx}\left(A \frac{d(\phi'\sqrt{F_{X_s}A})}{dx}\right) \right|_{x=x(z)}.
\label{u(z)}
\ee


\subsection{Deformation procedure}\label{deform-proc}


The deformation procedure is a practical method used to formulate novel analytical scalar field models~\cite{Bazeia:2002xg}. That method consists in to connect an analytical system $V(\phi)$ with another one $V(\chi)$ having unknown solutions, through a deforming function $f(\chi)$.  The deformed system describing the scalar field $\chi(x, t)$ is given by $\mathcal{L}=\mathcal{L}(\chi,Y)$ analogous to the non-standard Lagrangian density \eqref{lagran}. In this situation, the deformation procedure follows the steps proposed in Ref.~\cite{DMGD}. 

Given a known system with analytical solutions, the first-order equation may be written as a function of the field itself: $\phi'=R(\phi)$.  Similarly, the deformed model $V(\chi)$ presents $\chi'=S(\chi)$. The deformation function connects these two models through the transformation $\phi \rightarrow f(\chi)$, then
\be\label{SR}
S(\chi)=\frac{R(\phi \rightarrow f(\chi))}{f_{\chi}}.
\ee
Since $\phi(x)$ is the known solution of the first model ${\cal L}(\phi,X)$, then the inverse function $\chi(x) = f^{-1}(\phi(x))$  is the static solution of the second model ${\cal L}(\chi,Y)$.

\begin{figure}%
\centering
\includegraphics[scale=0.4]{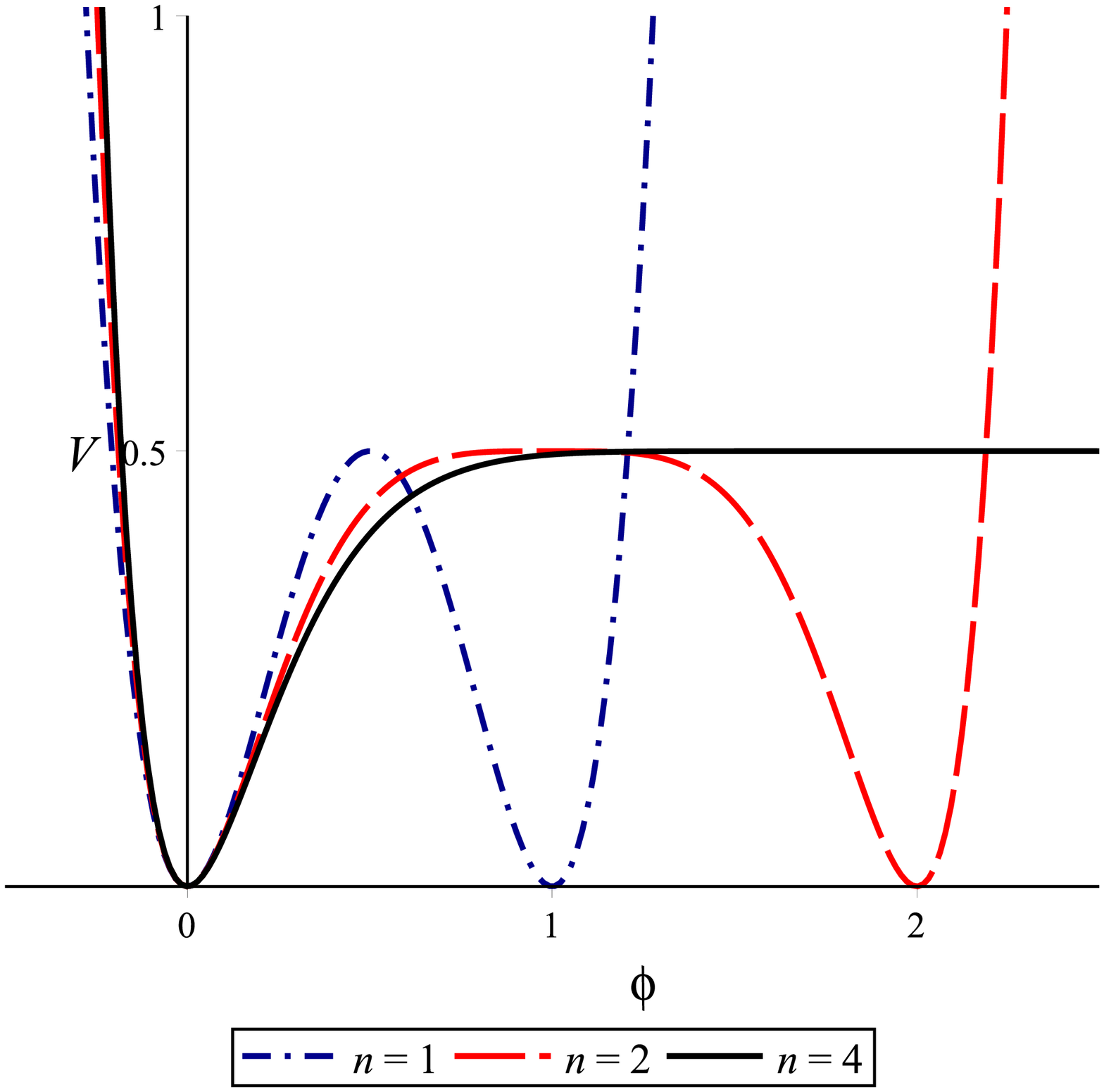}
\includegraphics[scale=0.4]{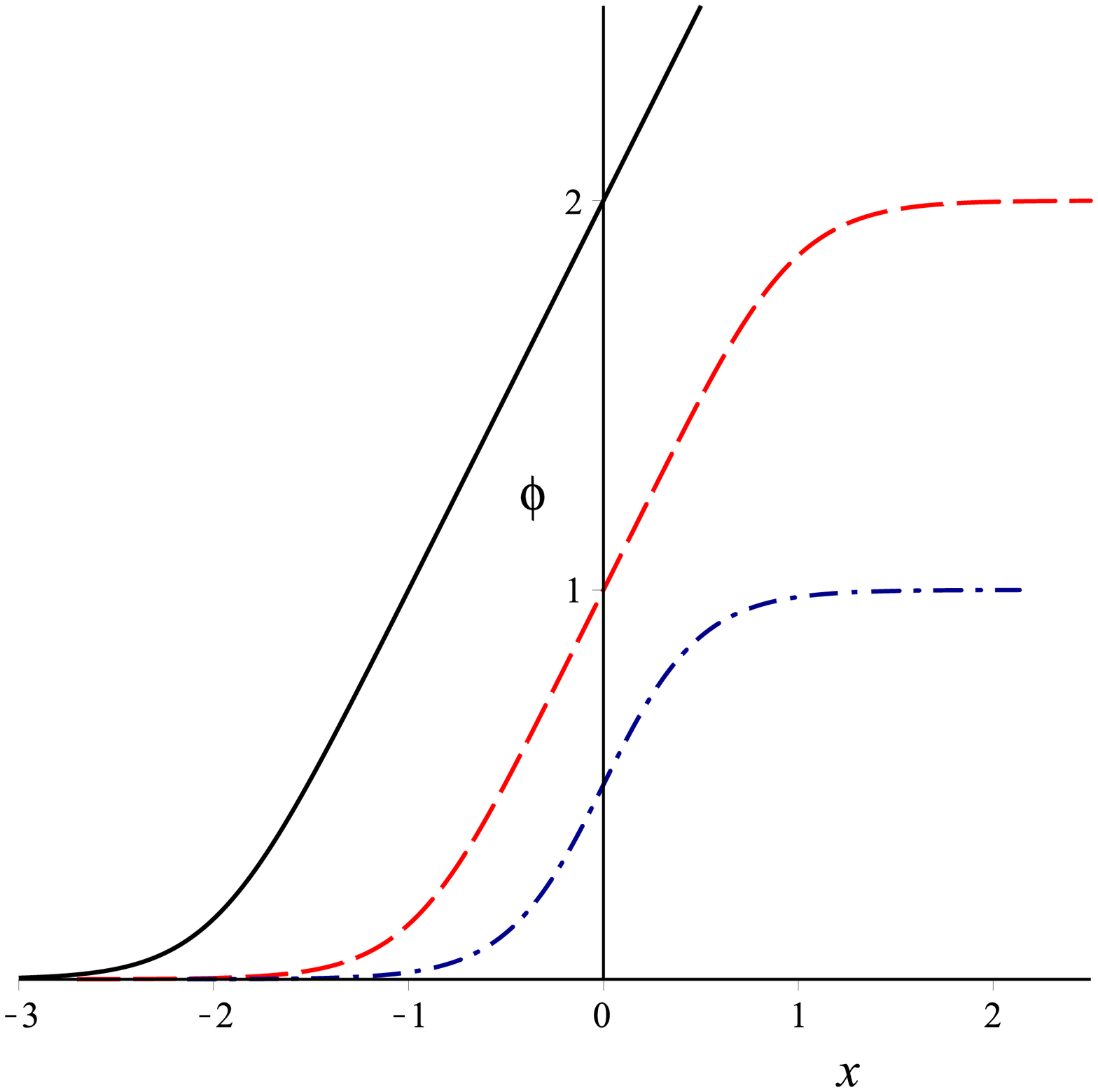}
\caption{The general $n-$Starobinsky model \eqref{Starobinsky} and the kink solutions, for $n=1,2,4$, represented  respectively by dashed-dot (blue), dashed (red) and solid (black) lines.}
\label{fig1}
\end{figure}

\section{Standard dynamics} \label{sec-2}
The standard theory is obtained by choosing $F(X)=X$, and the first-order formalism reduces to Bogomol'nyi-Prasad-Sommerfield (BPS) method \cite{bogomol}. One of the most well-known theories is the $\phi^4$ scalar field model  which is given by
\be
V(\phi)=\frac{1}{2}\left(1-\phi^2\right)^2.
\label{pnormal}
\ee
The static solutions are $\phi(x)=\pm \tanh(x)$, which have energy density $\rho(x)=\sech^4(x)$ and stability potential of P\"oschl-Teller type, $U(x)= 4-6\,\sech^2(x)$,  with one zero mode and one excited state \cite{teller}. Besides that, several other scalar field models of this kind with polynomial interactions are described in Ref.~\cite{bazeiaLeon} and references therein. 

\subsection{The  $n-$Starobinsky model}

Another model we suggest to analyze in this paper is the $\beta$-Starobinsky, proposed by the authors in Ref.~\cite{epjp-2021}, which is
\be
V(\phi)=\frac{V_0}{2}\left(1-\left(1-\sqrt{\frac23}\beta\frac{\phi}{M_{Pl}}\right)^{\frac1\beta}\right)^2.
\ee
The $\beta$-Starobinsky potential can be rewritten in terms of dimensionless quantities, $ V/V_0 \rightarrow V_n$ and  $\frac{\phi}{2\sqrt{6}M_{Pl}} \rightarrow \phi$, then
\be
V_{n}(\phi)=\frac12\left(1-\left(1-\frac2n\phi\right)^{2n}\right)^2,
\label{Starobinsky}
\ee 
where we have replaced the parameter $\beta$ to $\beta=1/2n$, with $n$ being a positive integer. In this situation, the potential presents a topological sector for each value of $n$. The minima are localized at $\phi_{min}=0,n$ and the maxima at $\phi_{max}=n/2$, as can be seen in Figure~\ref{fig1}. Employing the BPS-formalism, we note that
\be
W(\phi)=\phi+\frac{n}{2(2n+1)}\left(1-\frac2n\phi\right)^{2n+1}.
\ee
Then, we can calculate the energy of the topological solutions using the simplest form
\be
E_n=W(n)-W(0)=\frac{2n^2}{2n+1}.
\ee
This way, the energies of the kink-like defects are known, without calculating the solutions explicitly.

Furthermore, the expansion for $n>>1$ gives
\be
\left(1-\frac2n\phi\right)^{2n}=\e^{-4\phi}-\frac4n\phi^2\e^{-4\phi}+{\cal O}\left(\frac{1}{n^2}\right),
\ee
so that at the limit $n\rightarrow \infty$, the original Starobinsky model is recovered, as indicated on the first panel of Figure~\ref{fig1}.

\begin{figure}%
\centering
\includegraphics[scale=0.22]{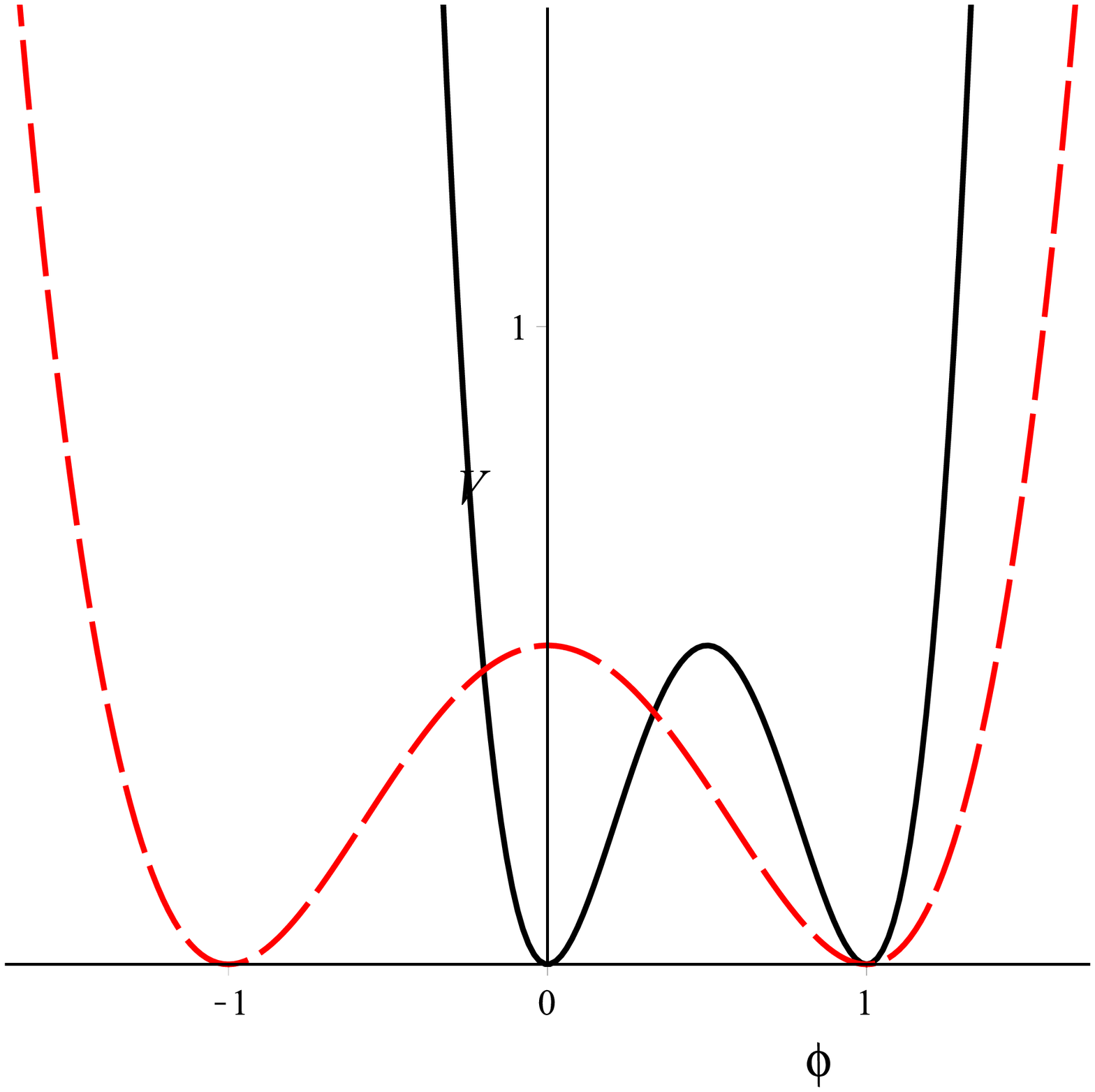}
\includegraphics[scale=0.22]{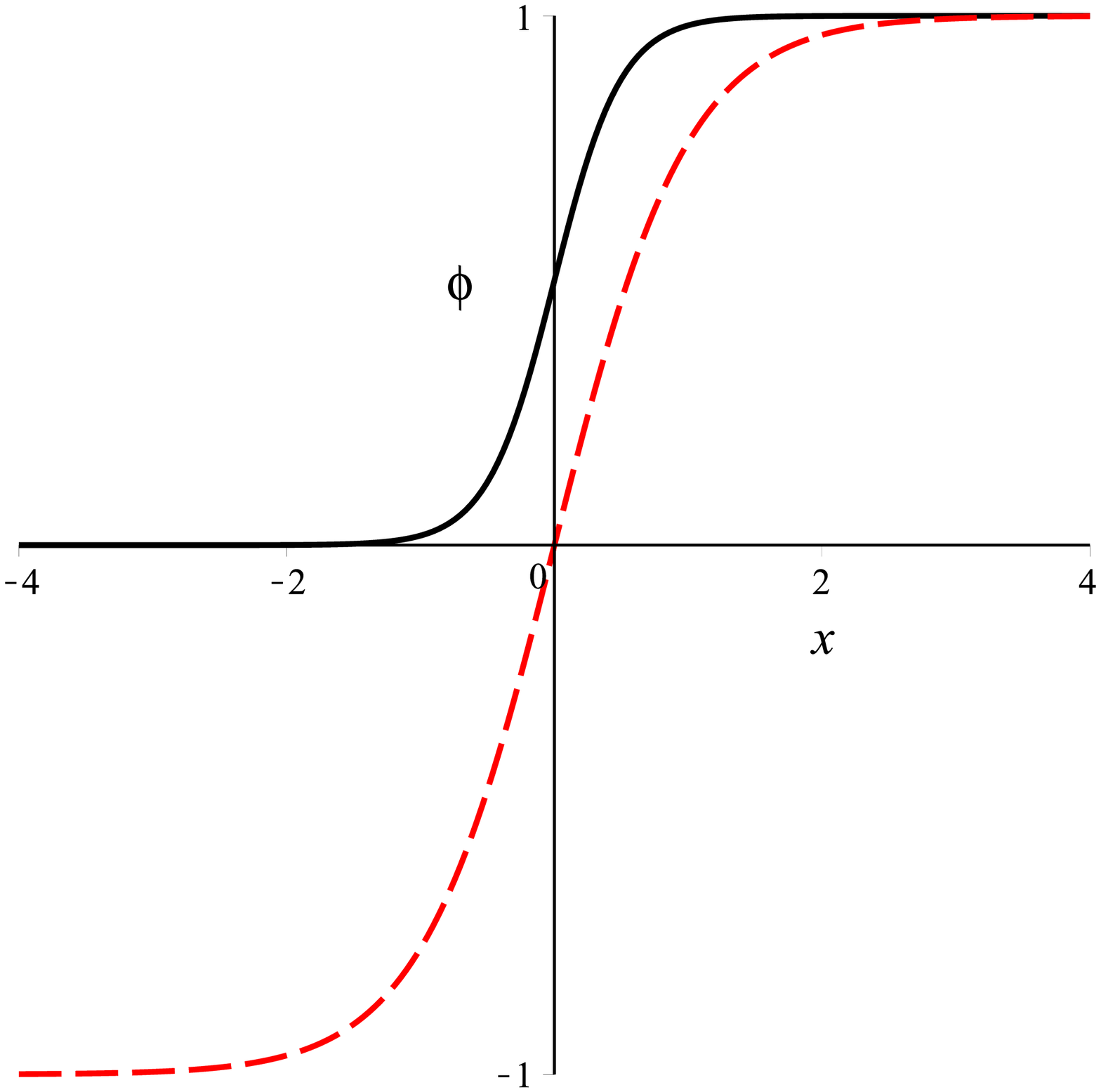}
\includegraphics[scale=0.22]{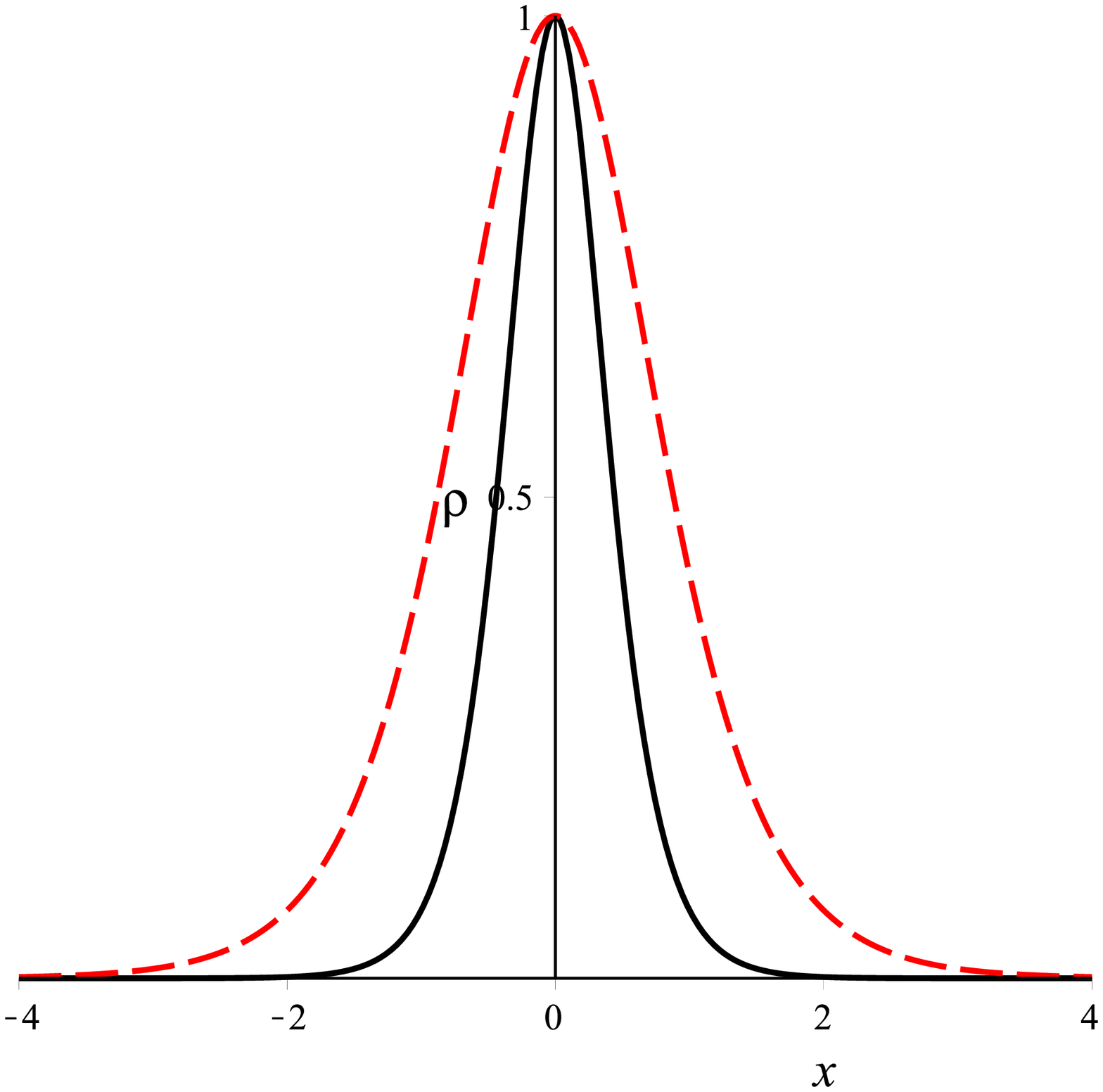}
\includegraphics[scale=0.22]{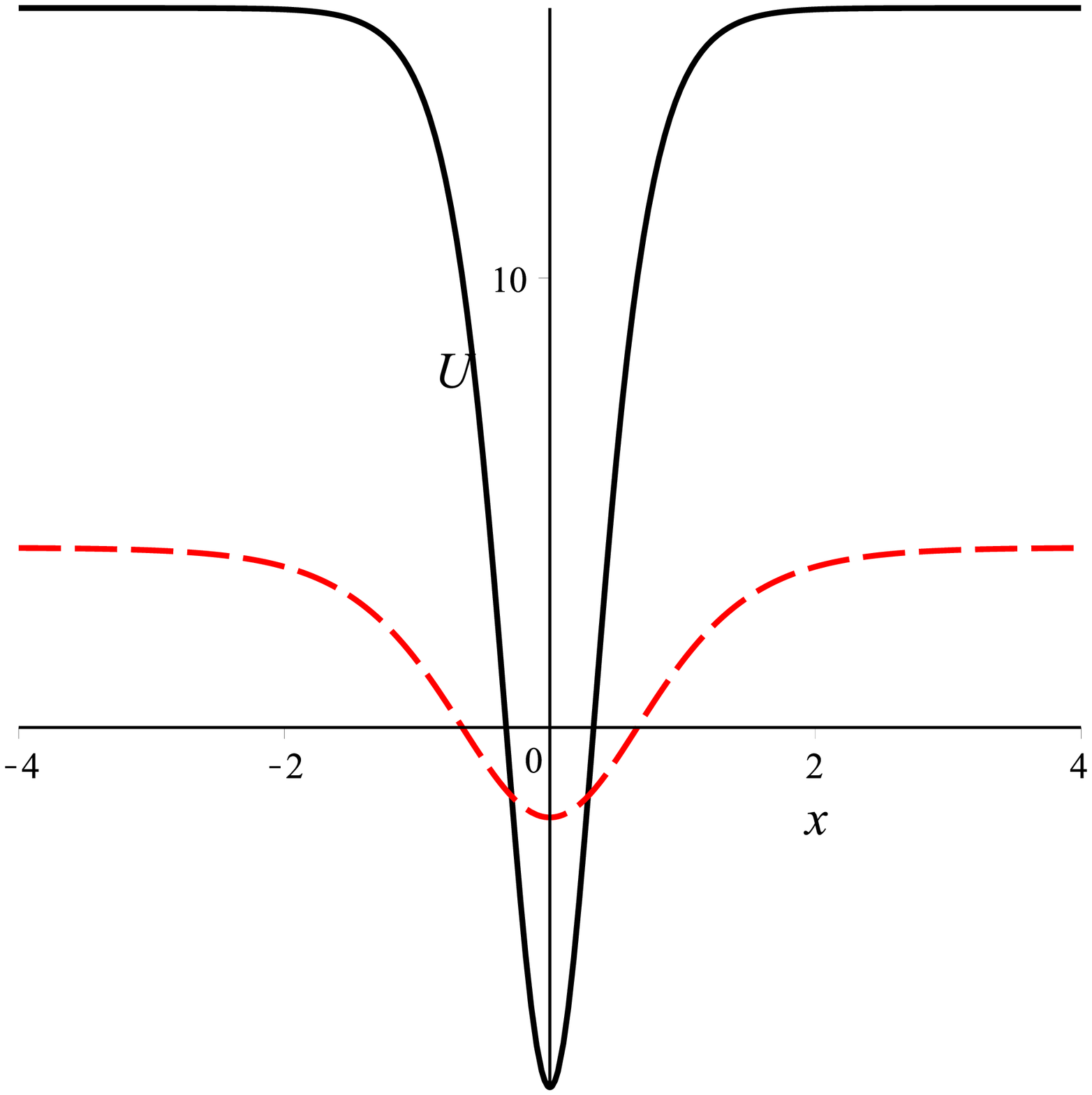}
\caption{Comparison between the $n-$Starobinsky model (with $n=1$) and the $\phi^4$ model, represented  by solid (black) and  dashed (red)  lines, respectively. The potentials \eqref{pnormal} and \eqref{v1sta} (first panel), their kink solutions (second panel), the energy densities (third panel), and stability potentials (fourth panel).}
\label{fig2}
\end{figure}

Considering the particular case $n=1$, the $n$-Starobinsky potential reads
\be
V_1(\phi)=8\phi^2\left(1-\phi\right)^2,
\label{v1sta}
\ee
which is like a $\phi^4$-model with an additional term $\phi^3$. The first-order equation is $\phi'=4\phi(1-\phi)$, and the solutions are
\be
\label{sol1staro}
\phi(x)=\frac{1 \pm \tanh(2x)}{2}.
\ee
The energy density is $\rho(x)=\sech^4(2x)$, that furnishes the energy $E=2/3$. The stability potential is
\be
U(x)=4(4-6\,\sech^2(2x)),
\ee
which has two bound states with eigenvalues $\omega_0=0$ and $\omega_1^2=12$. Figure~\ref{fig2} shows a parallel between the $\phi^4$-model and the $n-$Starobinsky on this particular case $n=1$.

\begin{figure}%
\centering
\includegraphics[scale=0.4]{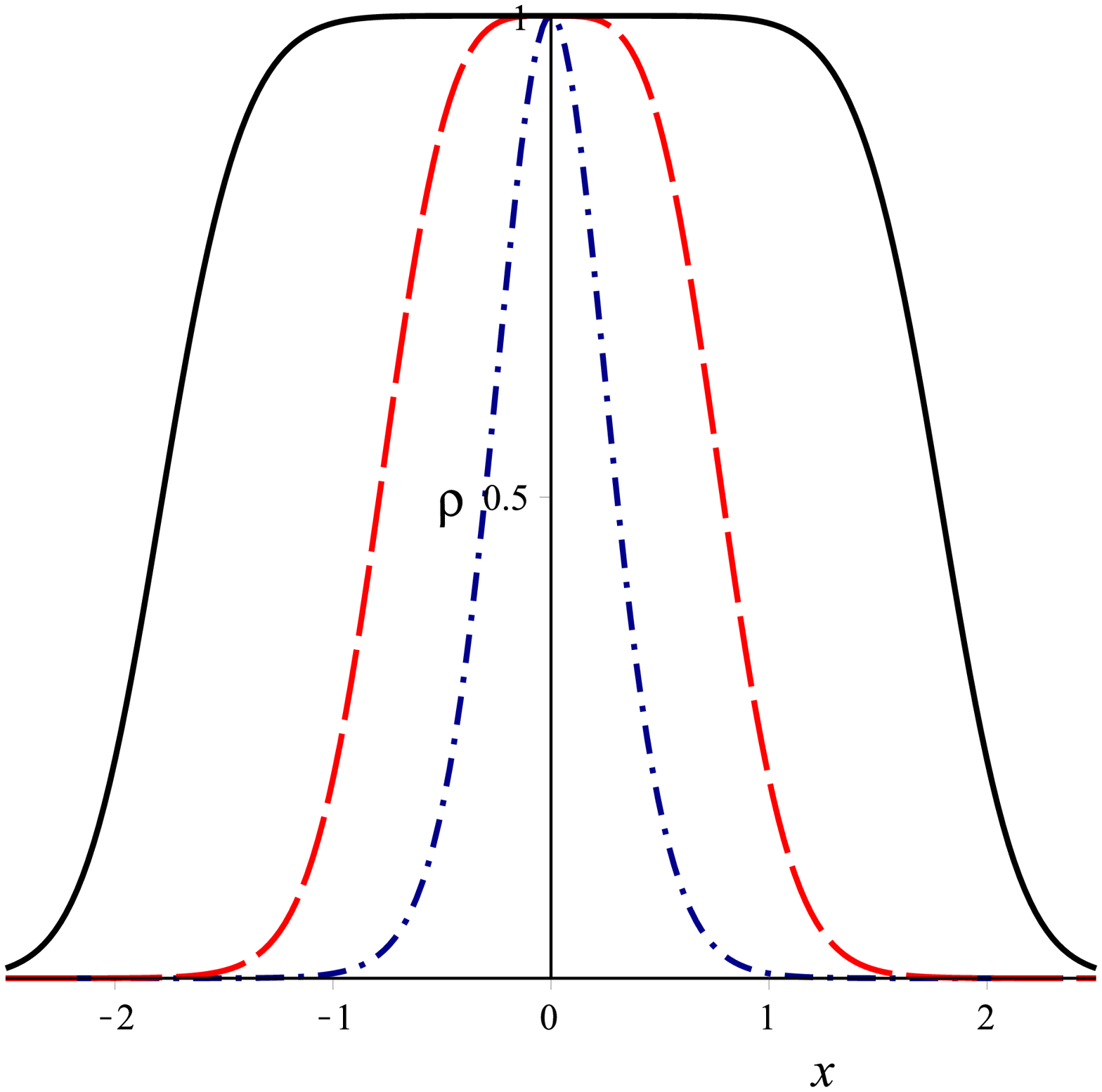}
\includegraphics[scale=0.4]{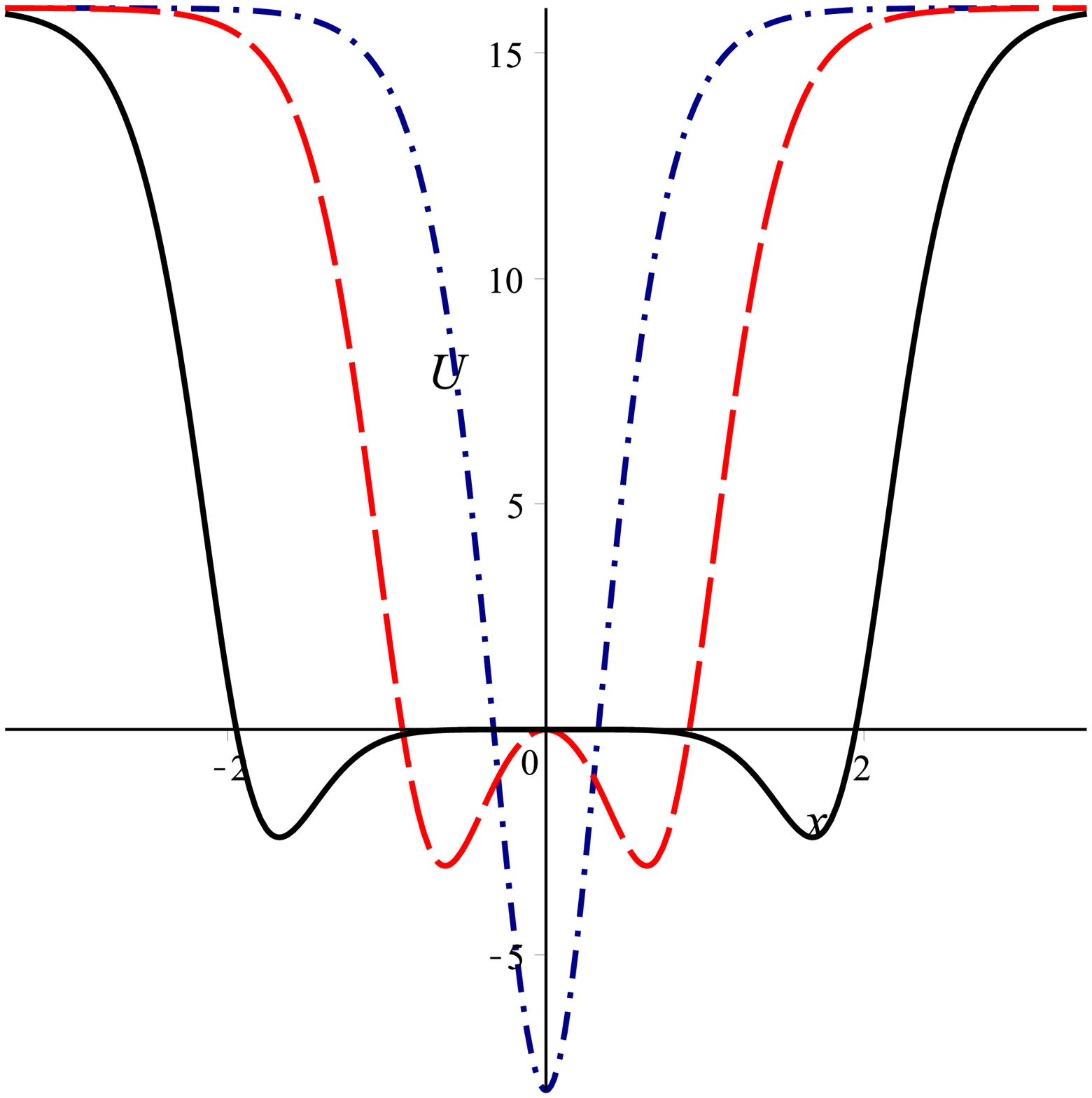}
\caption{The energy densities and stability potentials of the $n-$Starobinsky solutions, considering the same values of $n$ used in Fig.~\ref{fig1}.}
\label{fig1a}
\end{figure}

The $n-$Starobinsky general case can also be obtained through the deformation procedure. We select a deforming function that leads $V_1(\phi)\rightarrow V_n(\chi)$, which is
\be
\label{funcstaro}
f(\chi)=\frac12\left[1-\tanh\left(n\left(1-2\chi/n\right) {F_1}\right)\right].
\ee  
The functions $F_1\equiv{F_1}(1,1/2n;1+1/2n;\left(1-2\chi/n\right)^{2n})$ are ordinary hypergeometric functions. Then,
\be
W_{\chi}=\frac{4f(1-f)}{f_\chi}
\ee
becomes
\be
W_{\chi}=\left(\frac{d}{du}\left\{u \; {F_1}(1,1/2n;1+1/2n;u^{2n})\right\}\right)^{-1},
\ee
where we have called $u=\left(1-2\chi/n\right)$. Using the relation
\be
u\; {F_1}(1,1/2n;1+1/2n;u^{2n})=\int\frac{1}{1-u^{2n}}du,
\ee
the $n-$Starobinsky model is obtained
\be
W_{\chi}=1-\left(1-2\chi/n\right)^{2n},
\ee
and the kink solutions are given through the transcendental equation
\be
\label{hyptranscen}
\left(1-\frac2n\chi\right)\; {F_1}\left(1,\frac1{2n};1+\frac1{2n};\left(1-\frac2n\chi\right)^{2n}\right)= - \frac2n x,
\ee
which can also be written in terms of Lerch's transcendent $\Phi\left(u^{2n},1,1/2n\right)$, that is
\be
\label{lerchitranscen}
\left(1-\frac2n\chi\right)\Phi\left(\left(1-\frac2n\chi\right)^{2n},1,\frac1{2n}\right)=-4x.
\ee
These solutions are shown on the second panel of Figure~\ref{fig1} for some values of $n$. The antikinks are obtained changing $x\rightarrow -x$. The energy densities and stability potentials are represented in Fig.~\ref{fig1a}.

\section{Dirac-Born-Infeld dynamics} \label{sec-3}
 
Here, we propose to alter the standard Lagrangian by selecting a scalar field theory in which the kinetic term is non-canonical, according to Dirac-Born-Infeld dynamics
\be
\label{FBI}
F(X)=-\alpha^2\left(\sqrt{1-2X/\alpha^2}-1\right),
\ee
where $\alpha$ is a non-null real parameter. Taking the expansion for $|\alpha|>>1$, one gets  
\be
F(X)=X+\frac{1}{2\alpha^2}X^2+\frac{1}{2\alpha^4}X^3+{\cal O}\left(\frac{1}{\alpha^6}\right).
\ee
As can be seen, the parameter  $\alpha$ controls high-order powers of the field derivatives. For $|\alpha| \rightarrow \infty$, the ordinary kinetic term is recovered. 

Considering the first-order description previously discussed, the static field derivative is written in terms of the function $W_{\phi}$ 
\be
\label{fow}
\phi'=\frac{W_{\phi}}{\sqrt{1-W_{\phi}^2/\alpha^2}}.
\ee
It implies that $F_X=\sqrt{1-W_\phi^2/\alpha^2}$.  Furthermore, another equation related to \eqref{fow} is obtained taking $x\rightarrow -x$.

The potential, $V=F+\phi'^2 F_X$, in the context of the DBI kinetics may also be written as $V=\alpha^2(1-F_X)$. That in terms of $W$ is 
\be
\label{potw}
V(\phi)=\alpha^2\left(1-\sqrt{1-W_{\phi}^2/\alpha^2}\right).
\ee

In this work, we are interested in constructing potentials supporting topological structures; such that the equations of motion can be analytically soluble employing first-order equations. The standard Lagrangian modification leads to a square root on the potential, which introduces some restrictions on the scalar field. In the limit, $|\alpha|>> 1$, the scalar field acquires a usual dynamics, and the standard Lagrangian is recovered
\be
{\cal L}(\phi,X)=X-\frac{1}{2}W_{\phi}^2 + \frac{1}{2\alpha^2}\left(X^2-\frac14 W_\phi^4\right) +
O\left(\frac{1}{\alpha^4}\right),
\ee
ensuring an approach compatible with the original BPS description \cite{bogomol}.

The energy density \eqref{rho} becomes
\be
\label{rhofo}
\rho(x)=\frac{\phi'^2}{\sqrt{1+\phi'^2/\alpha^2}}=\frac{W_{\phi}^2}{\sqrt{1-W_{\phi}^2/\alpha^2}}.
\ee
The linear stability potential, Eq.~\eqref{u(z)}, becomes
\be
U(z)=\left. \frac{1}{\phi'}\frac{d}{dx}\left(F_{X}V_\phi \right) \right|_{x=x(z)},
\label{staby}
\ee
since in the DBI case, the static equation of motion is $(\phi'F_X)'=V_\phi$ and  $F_{X}=A$ .

Taking into account the first-order formalism, we have $F_{X} V_\phi=W_\phi W_{\phi\phi}$; consequently the stability potential reads
\be
U(z)= \left. W_{\phi\phi}^2+W_\phi W_{\phi\phi\phi} \,\right|_{x=x(z)}.
\ee
The Schr$\ddot{\o}$dinger-like equation \eqref{schrodingerlikequation} can be factorizable as $Q^{\dagger}Q \varepsilon_n=\omega_n^2\varepsilon_n$, with
\be
Q=-\frac{d}{dz}+W_{\phi\phi}  \:\:\: \:\:\: \mbox{and}   \:\:\: \:\:\:  Q^{\dagger}=\frac{d}{dz}+W_{\phi\phi},
\ee
since $\phi_z=W_{\phi}$. Therefore, the eigenvalues $\omega_n^2$ are positive, and the static solutions are linearly stable. Besides,
there is at least a zero mode given by
\be
\varepsilon_0(z)={\cal N}\phi_z= {\cal N} W_{\phi},
\ee 
where ${\cal N}$ is a normalization constant. 

The deformation procedure described in Sec.~\ref{sec-1} gives
\be
\label{DP}
V(\chi)= \alpha^2\left(1-\frac{1}{\sqrt{1+S(\chi)^2/\alpha^2}}\right).
\ee
The stability potential becomes
\be
U(z)=\left. \frac{1}{S(\chi)}\frac{d}{dx}\left(\frac{S(\chi)}{\left(1+S(\chi)^2/\alpha^2\right)^2}\frac{d S}{d\chi}\right)\right|_{x(z)}.
\ee

\subsection{Models relatives of polynomial potentials}

\begin{figure}%
\centering
\includegraphics[scale=0.22]{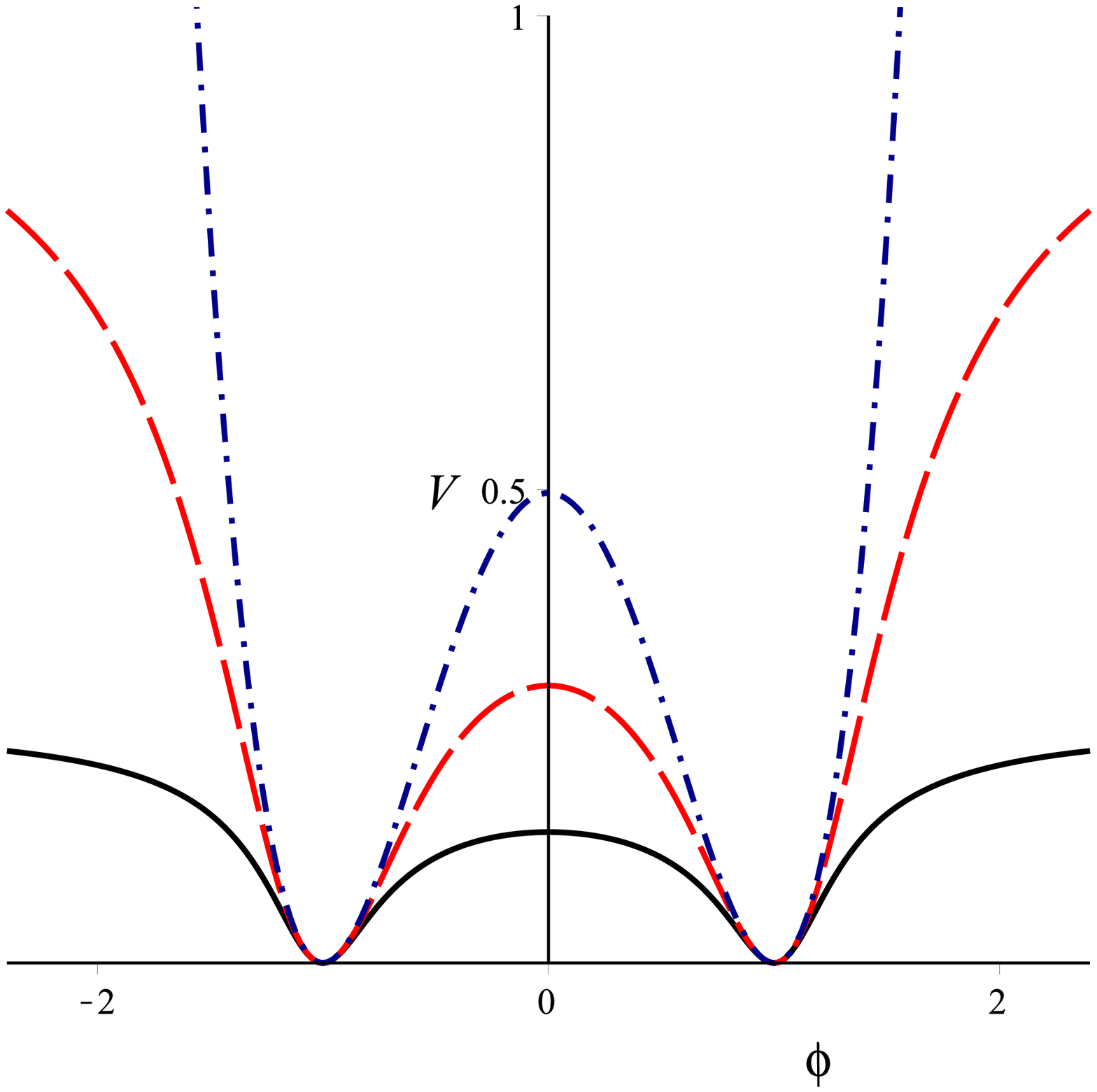}
\includegraphics[scale=0.22]{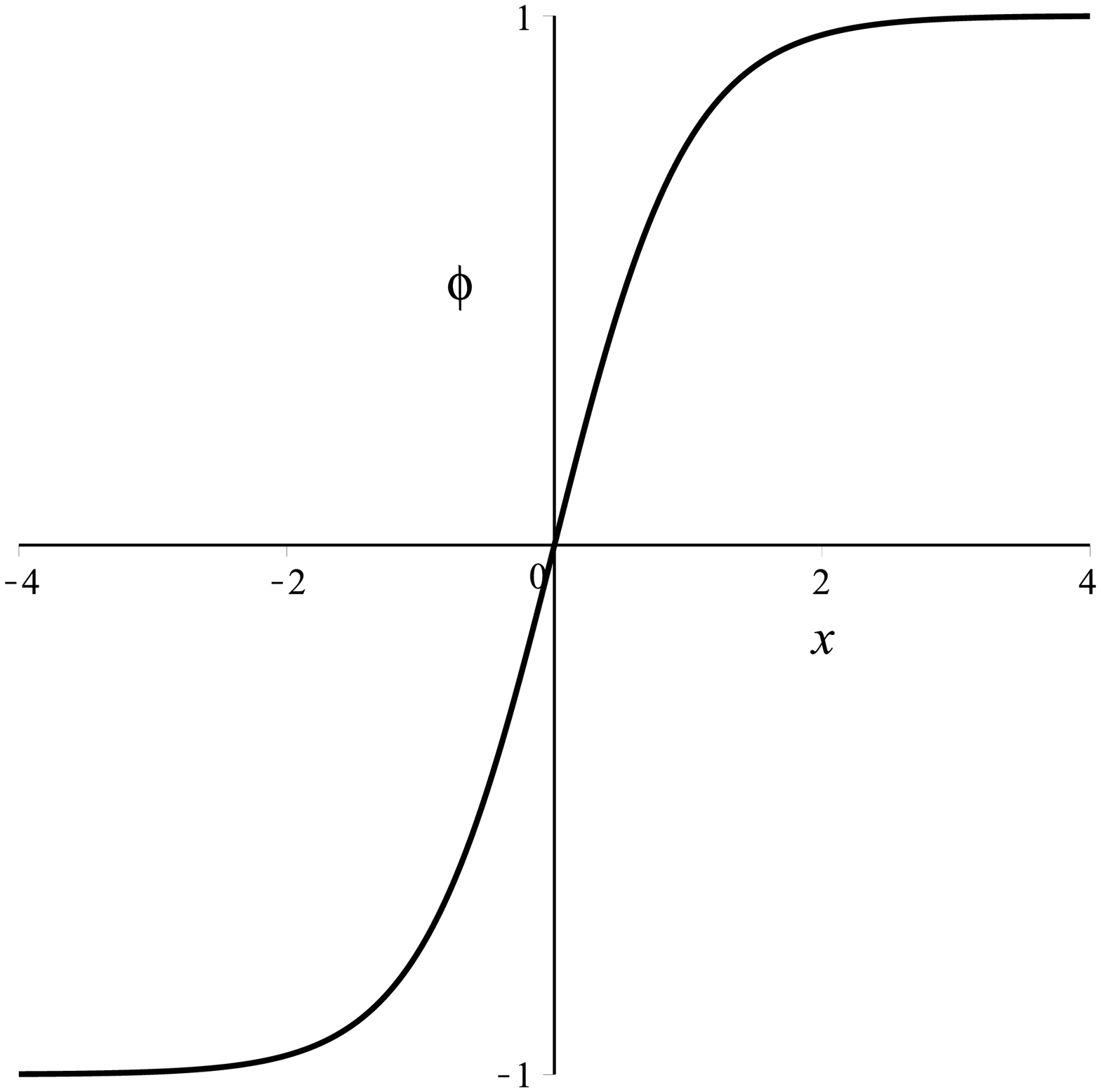}
\includegraphics[scale=0.22]{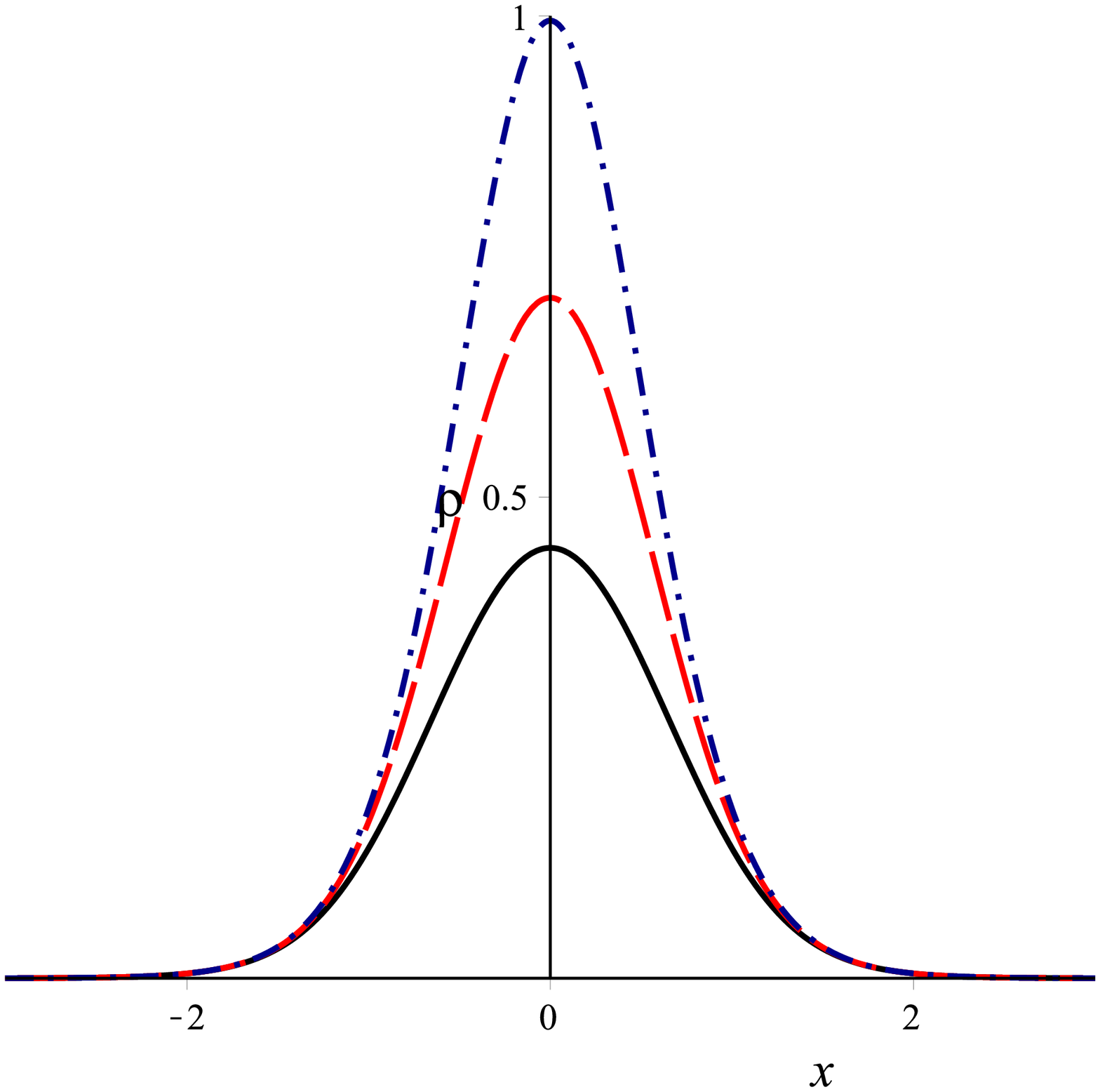}
\includegraphics[scale=0.22]{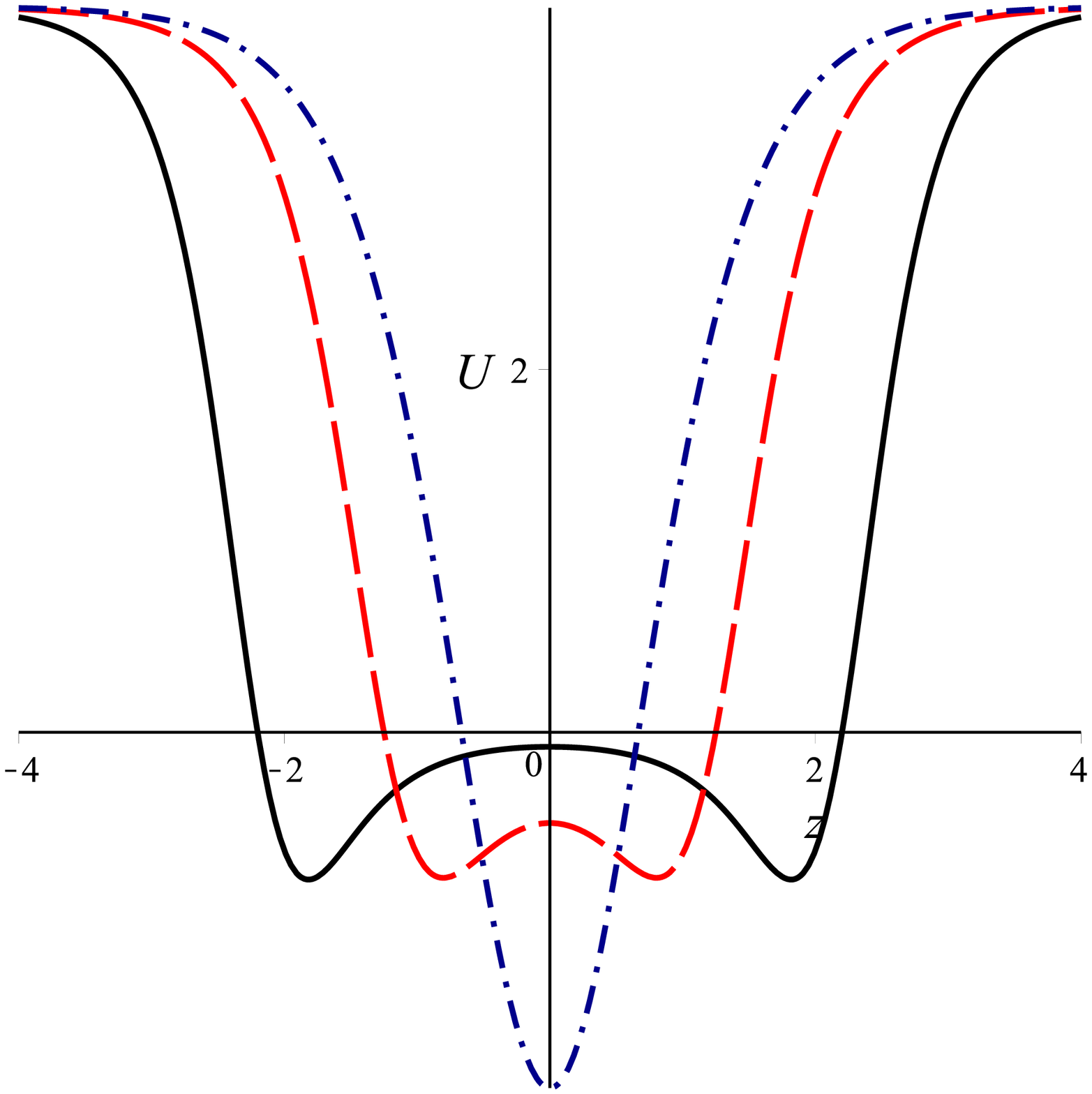}
\caption{The potential \eqref{pot1} (first panel), its kink solution (second panel), the energy density \eqref{rh1} (third panel), and the stability potential $U(z)$ calculated through \eqref{staby} (fourth panel). We have chosen  ${\alpha} = 0.5, 1, 10$, represented respectively by solid (black),  dashed (red), and  dashed-dot (blue)  lines.}
\label{plot1}
\end{figure}

In order to find novel analytical models relatives of polynomial potentials, we select a known model which is going to be deformed
\be
\label{pot1}
V(\phi)=\alpha^2\left(1-\frac{1}{\sqrt{1+\dfrac{1}{\alpha^2}(1-\phi^2)^2}}\right).
\ee
This potential is proposed in Ref.~\cite{Bazeia:2017mnc}, which shall be a deforming model. That is a non-negative potential with minima at $\bar{\phi}=\pm 1$ and maximum at the origin, as can be seen at the leftmost panel of Fig.~\ref{plot1}. The kink solution is known and given by $\phi(x)= \tanh(x)$, that is the same solution found for the standard $\phi^4$ model. However, the DBI kinetic modifies the energy density and linear stability, in a way that 
\be
\label{rh1}
\rho(x)=\frac{\sech^4(x)}{\sqrt{1+\frac{1}{\alpha^2}\sech^4(x)}},
\ee
and
\be
F_{X}V_\phi=-\frac{2\,\sech^2(x)\tanh(x)}{\left(1+\frac{1}{\alpha^2}\sech^4(x)\right)^2},
\ee
as represented on the rightmost panels of Fig.~\ref{plot1} for some values of $\alpha$. To obtain $U(z)$ is necessary to perform a numerical integration, since the change of variables $z(x)$ carries a non-analytical integral equation.

In addition, the first-order equation is $\phi'=1-\phi^2 \equiv R(\phi)$. Through the deformation procedure, the function $S(\chi)$ becomes
\be
S(\chi)=\frac{1-f^2}{f_{\chi}}.
\ee
The solutions of novel field models shall be found by $\chi(x) = f^{-1}(\tanh(x))$. The same expression for $S(\chi)$ is provided by the deformation function $g(\chi)=1/f(\chi)$. Such distinct deforming functions lead exactly to the same deformed potential and furnish solutions to distinct topological sectors of $V(\chi)$ \cite{familySine2009}.

\subsubsection{The modified polynomial models}
The novel deformed models shall be obtained using the deformation procedure presented in the last sections, where we will introduce deforming functions based on the ones presented in Ref.~\cite{Bazeia:2017mnc} and references therein. However, we are going to suggest the change of $\chi\rightarrow \chi(1+\chi/4\alpha^2)$.
\begin{figure}%
\centering
\includegraphics[scale=0.22]{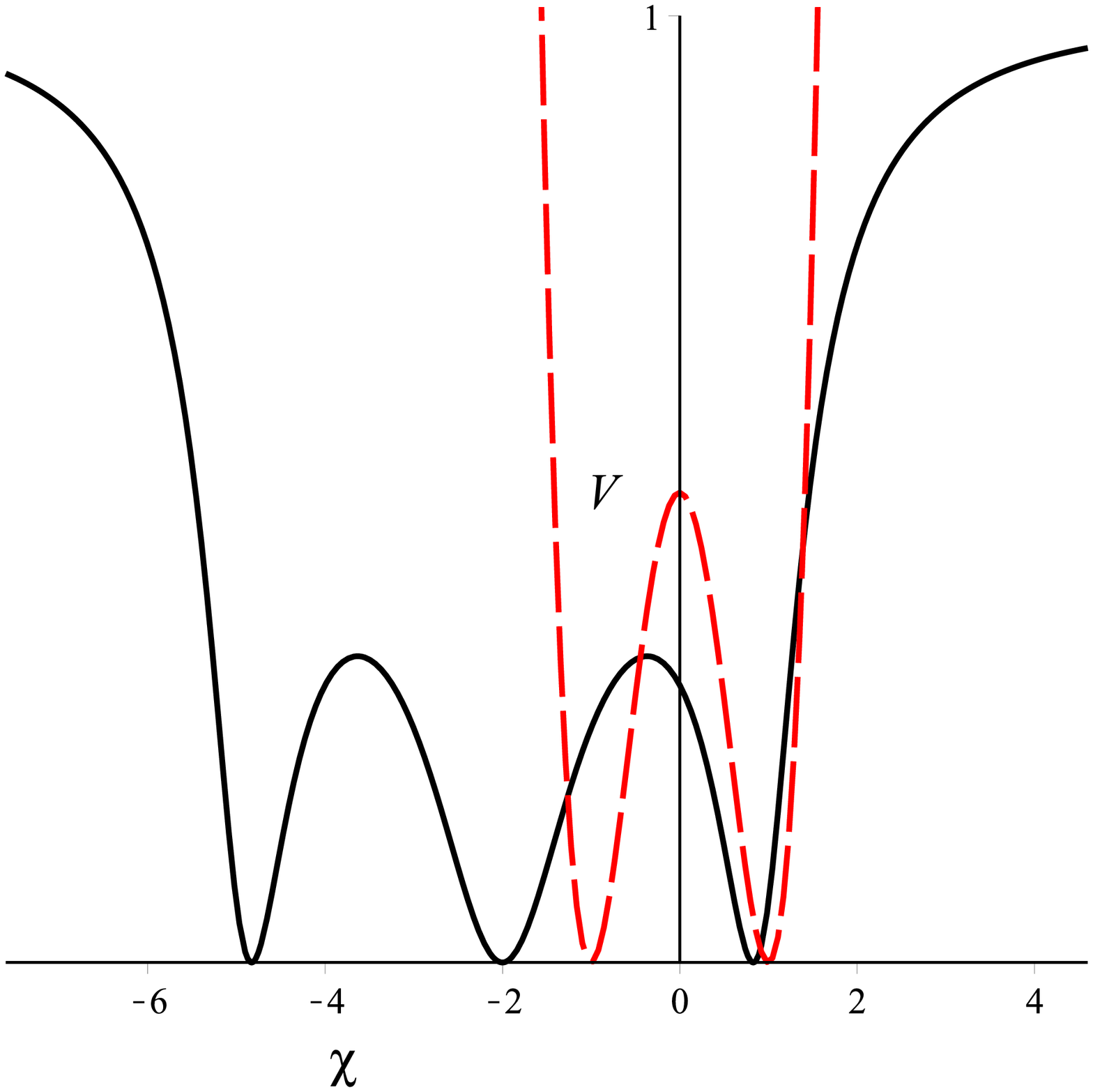}
\includegraphics[scale=0.22]{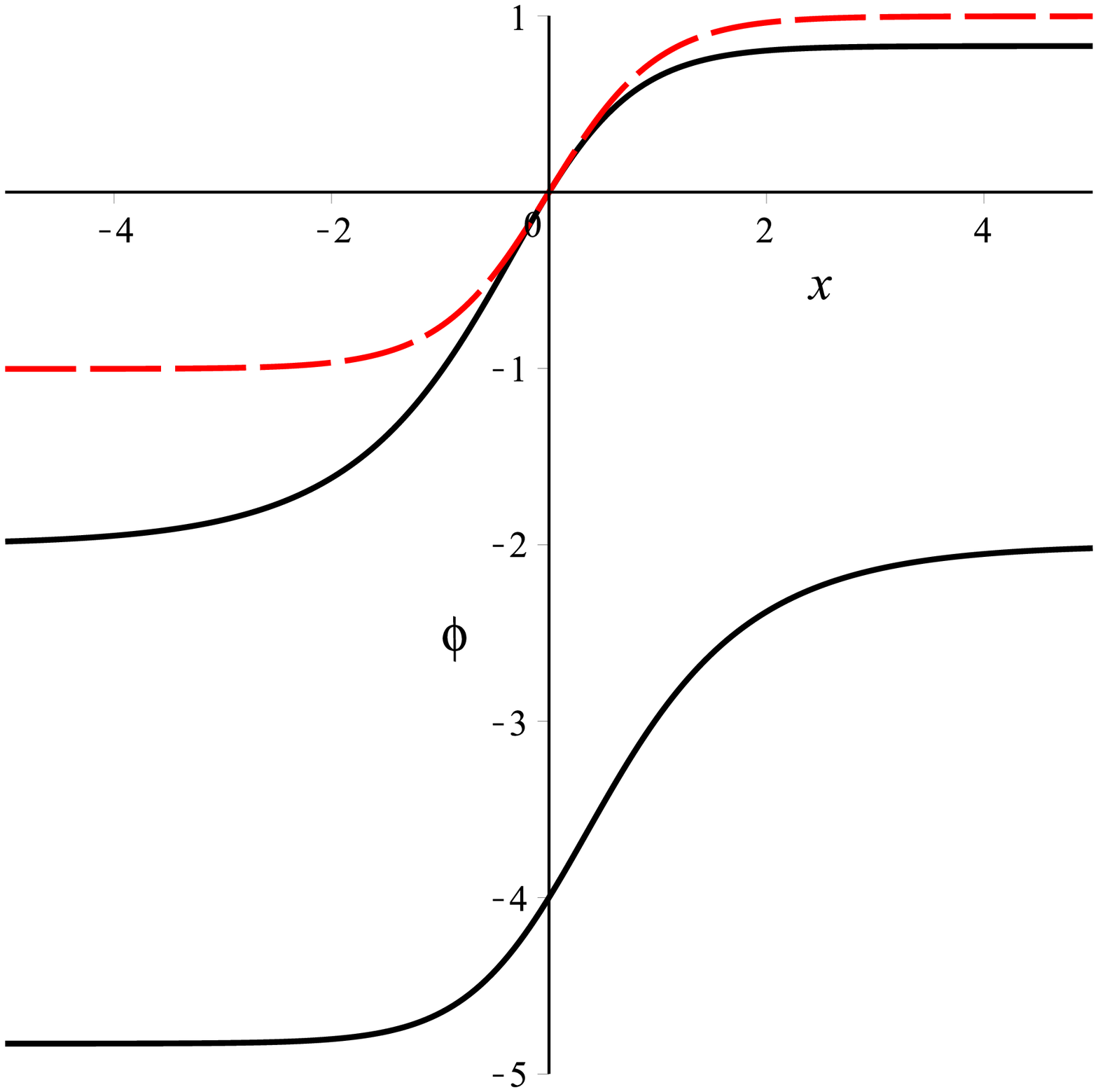}
\includegraphics[scale=0.22]{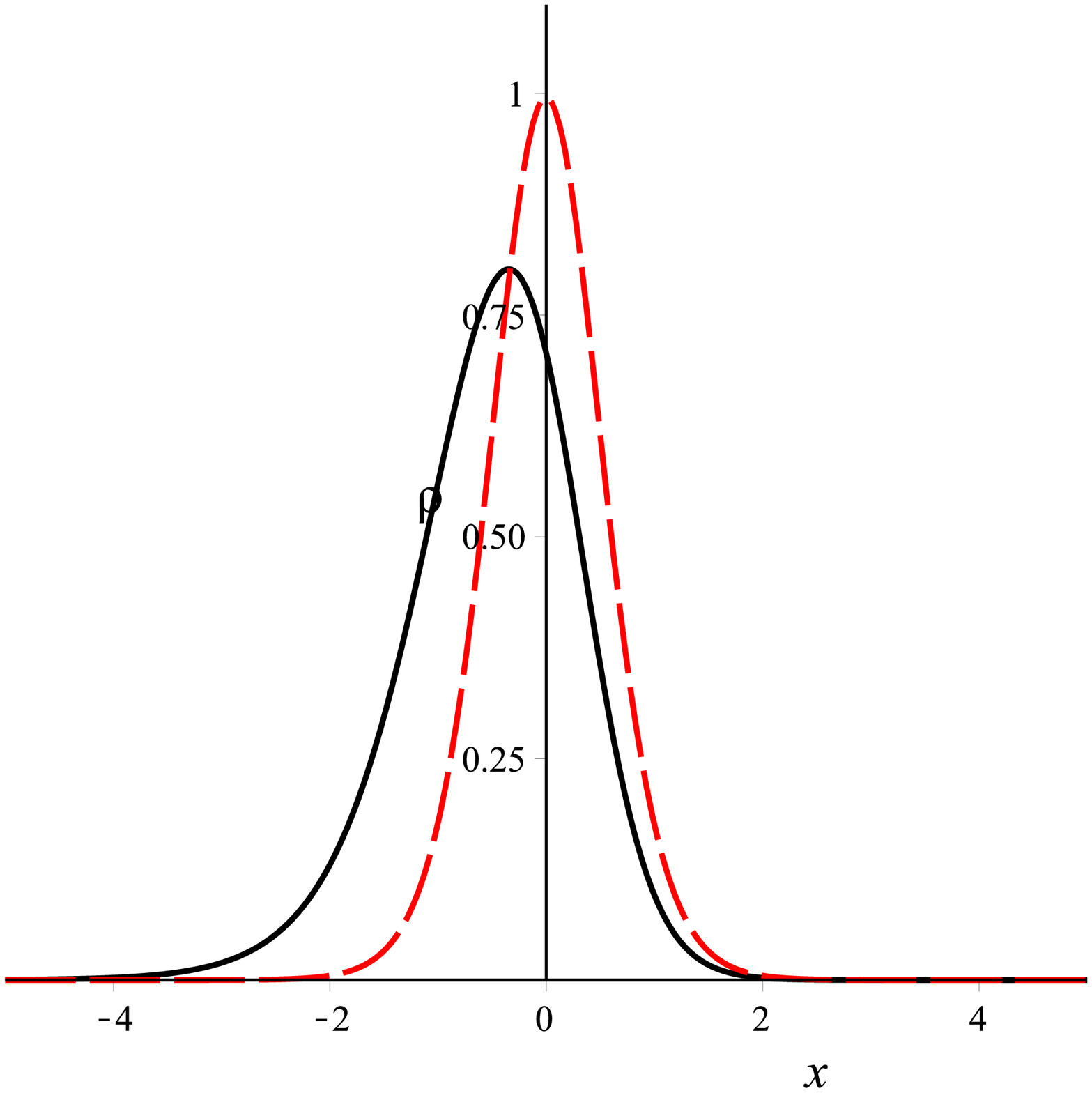}
\includegraphics[scale=0.22]{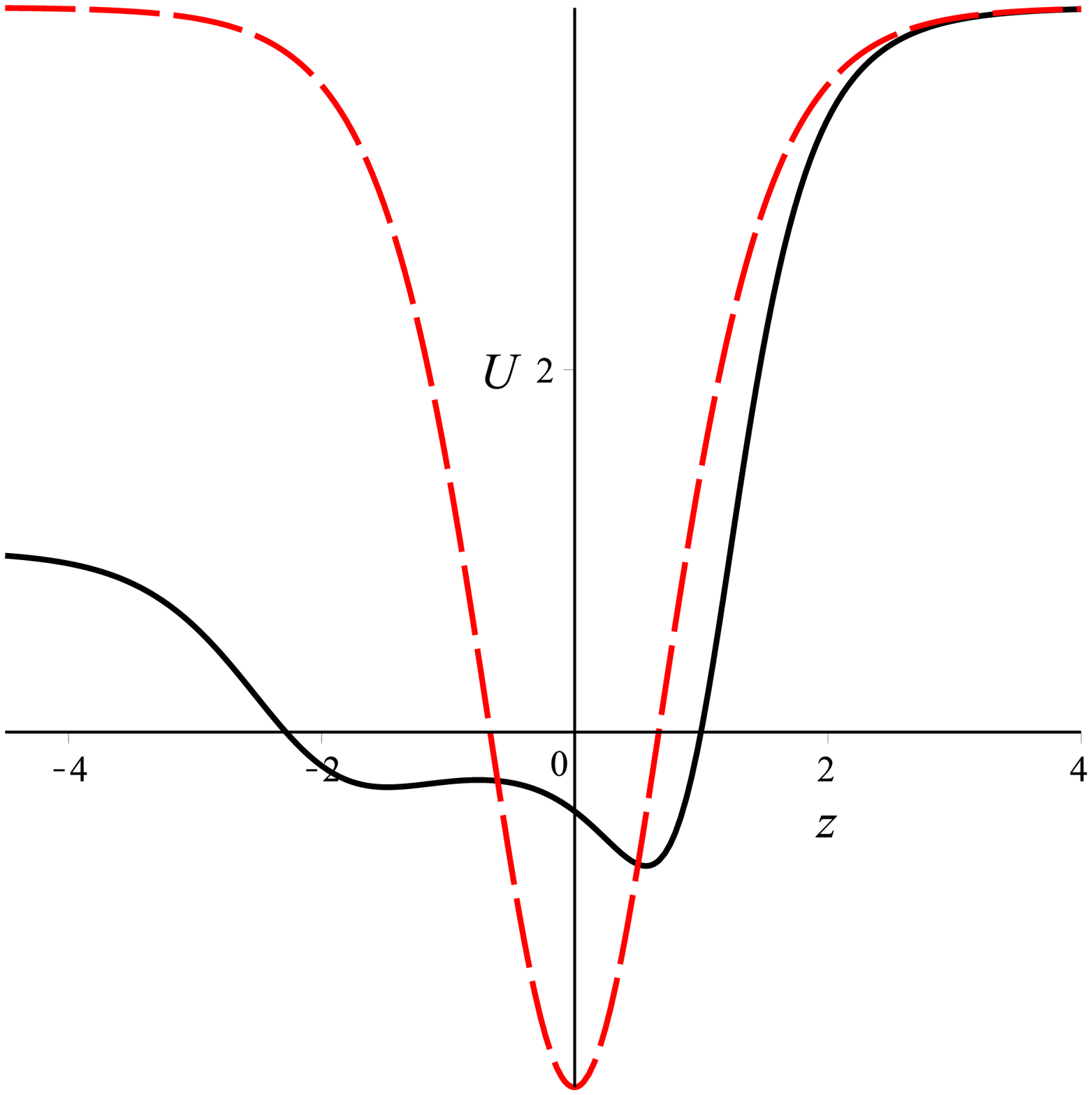}
\caption{The potential \eqref{V2}, the kink solutions \eqref{p2}, the energy density $\rho_{+}(x)$ given by Eq.~\eqref{dens2}, and the stability potential $U_{+}(z)$. For ${\alpha} = 1, 10$, represented respectively by solid (black) and  dashed (red) lines.}
\label{plot2}
\end{figure}

The first deformation function is $f_1(\chi)=\chi\left(1+\chi/4\alpha^2 \right)$, then
\be
\label{S1}
S_1(\chi)=\frac{1-\chi^2\left(1+{\chi}/{4\alpha^2} \right)^2}{1+{\chi}/{2\alpha^2}}.
\ee
The potential given by Eq.~\eqref{DP} becomes

\be
V = \alpha^2\left(1-\frac{|1+{\chi}/{2\alpha^2}|}{\sqrt{(1+\frac{\chi}{2\alpha^2})^2+\frac1{\alpha^2}\left(1-\chi^2\left(1+\frac{\chi}{4\alpha^2}\right)^2\right)^2}}\right) 
\label{V2}
\ee
which has minima at $\chi_{min}=2\alpha^2(-1\pm\sqrt{1\pm1/\alpha^2})$. For the particular case $\alpha=1$, there are three minima at the points $-2(1\pm\sqrt{2})$ and $-2$, and two maxima at $-2(1\pm\sqrt{2/3})$, constituting two topological sectors as  presented in the first graph of Fig.~\ref{plot2}. In the situation $\alpha=1$, the potential behaves like a $\phi^6$ model; but when $|\alpha|>>1$ the standard $\phi^4$ model is recovered.

The solutions are found by $\chi(x)=f_1^{-1}(\tanh(x))$, which are
\be
\label{p2}
\chi_{\pm\pm}(x)=2\alpha^2\left(-1\pm \sqrt{1\pm\frac{1}{\alpha^2}\tanh(x)}\right),
\ee
where $\chi_{++}$ is the kink solution corresponding to the right topological sector, and $\chi_{--}$ to the left one. Real value solutions require $|\alpha|\geq 1$. In the limit  $|\alpha|>>1$, these solutions become $\chi_{+\pm}(x)=\pm\tanh(x)$ and $\chi_{-\pm}(x)=-4\alpha^2\pm\tanh(x)$, in such way that $\chi_{-\pm}(x)$ is displaced to minus infinity, and the standard framework is recovered.

The energy density is
\be
\rho_\pm(x)=\frac{\sech^4(x)}{\left(1\pm\frac{1}{\alpha^2}\tanh(x)\right)\sqrt{1+\dfrac{\sech^4(x)}{\alpha^2\pm\tanh(x)}}},
\label{dens2}
\ee
$\rho_{+}(x)$ is related to the kink $\chi_{++}$ (anti-kink $\chi_{-+}$) of the right sector (left sector), and  $\rho_{-}(x)$ is equal to $\rho_{+}(x)$ mirrored. These quantities and the stability potential $U(z)$ are represented at the rightmost panels in Fig.~\ref{plot2}.

\begin{figure}%
\centering
\includegraphics[scale=0.22]{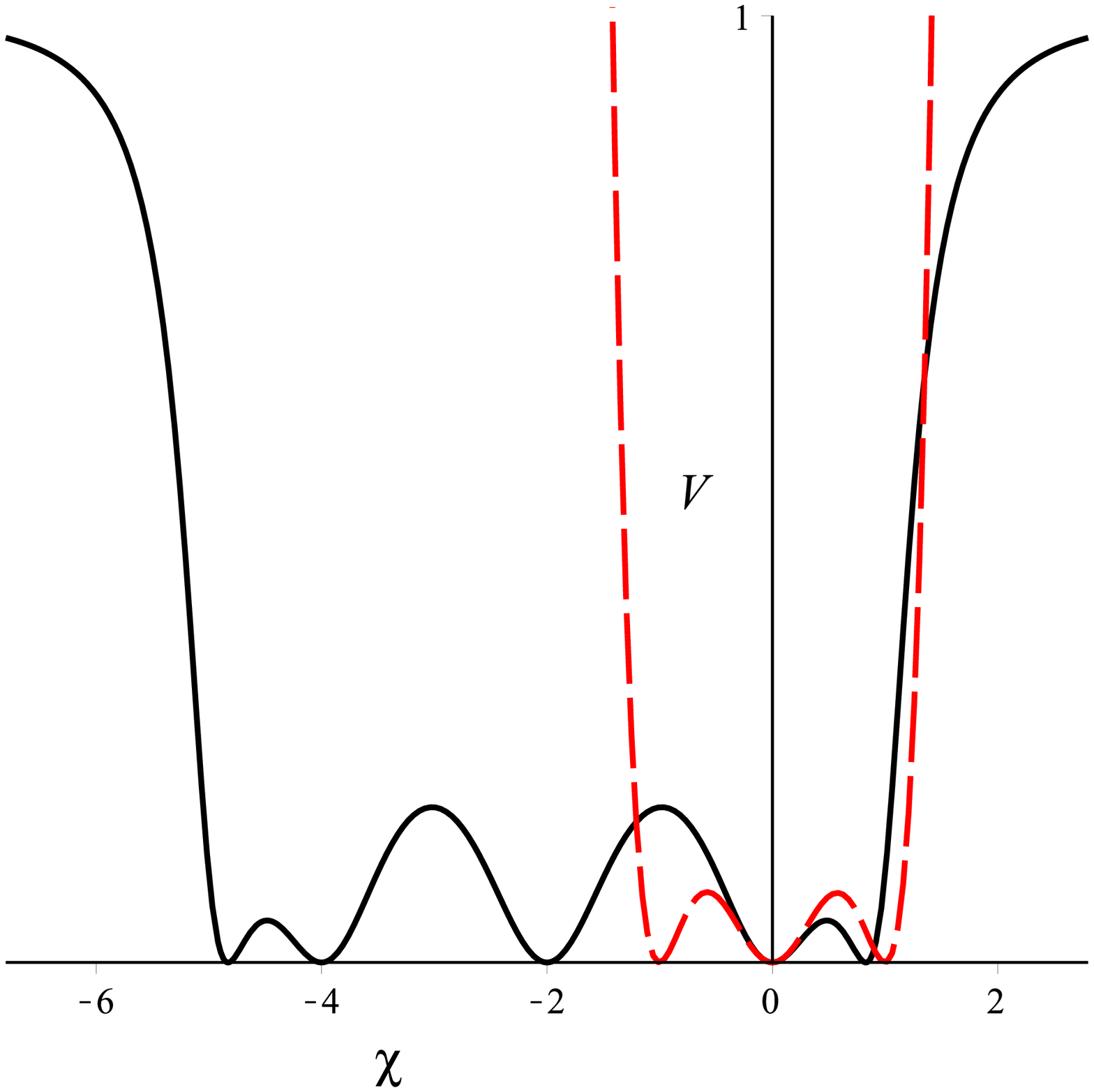}
\includegraphics[scale=0.22]{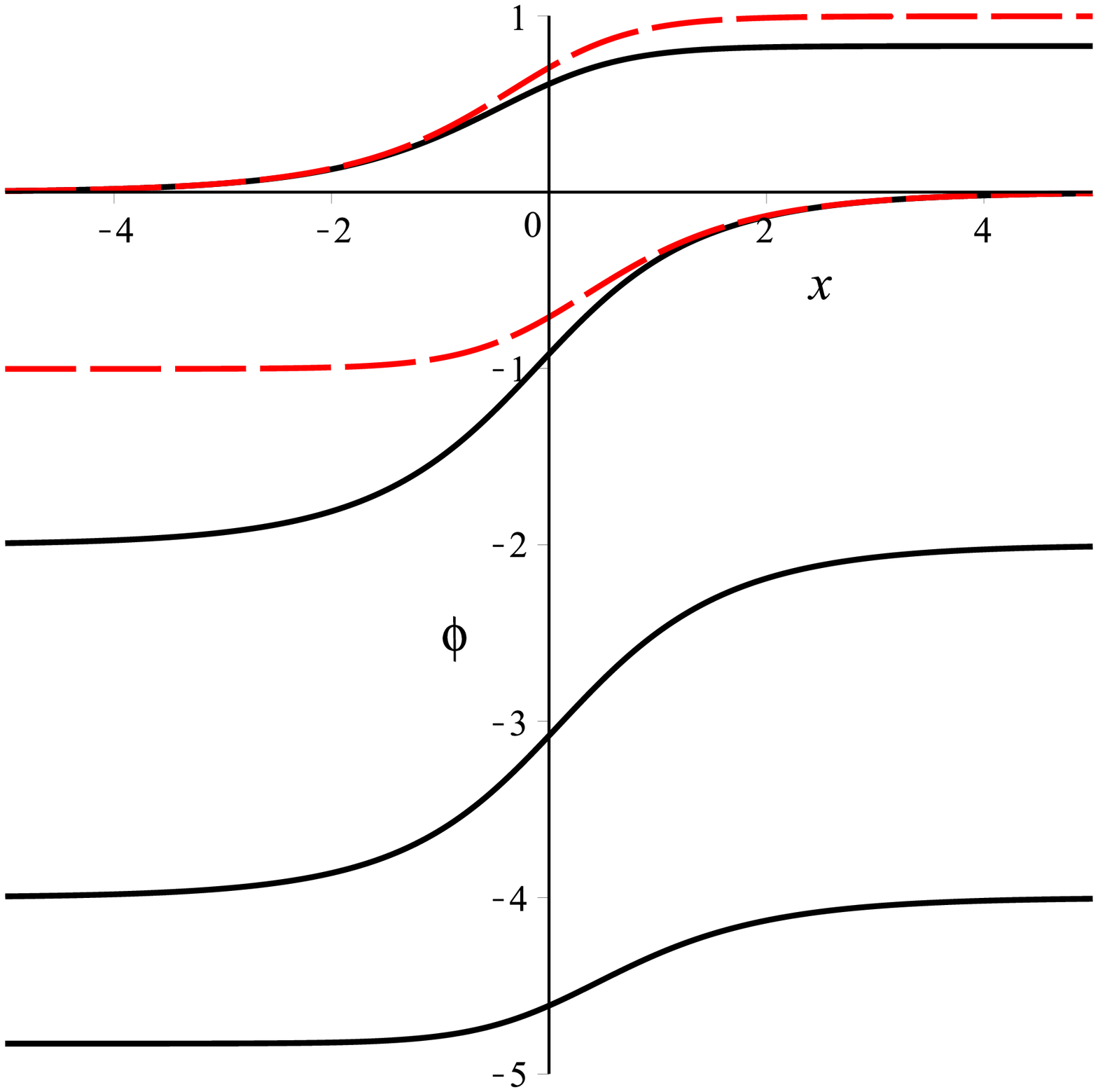}
\includegraphics[scale=0.22]{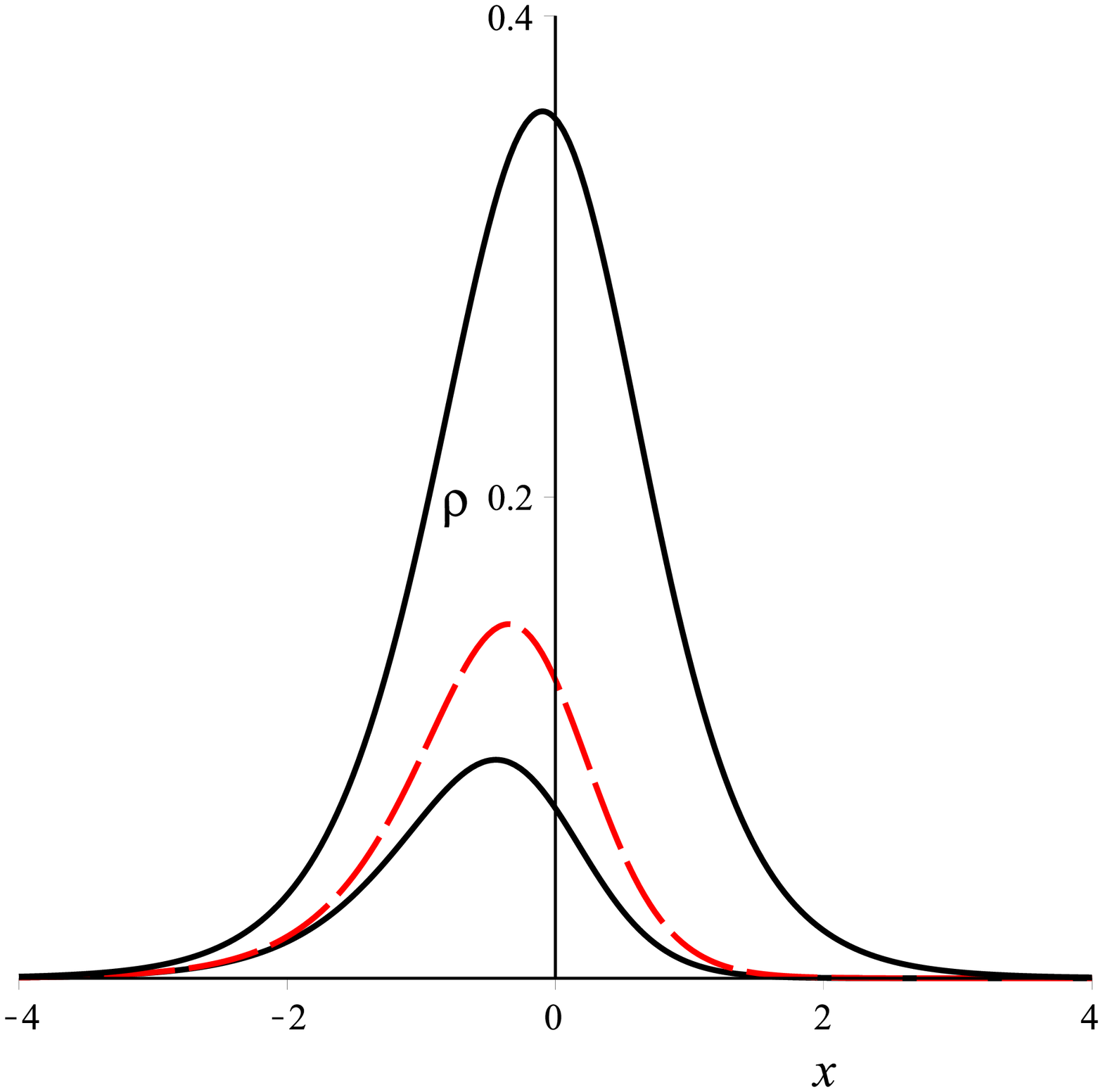}
\includegraphics[scale=0.22]{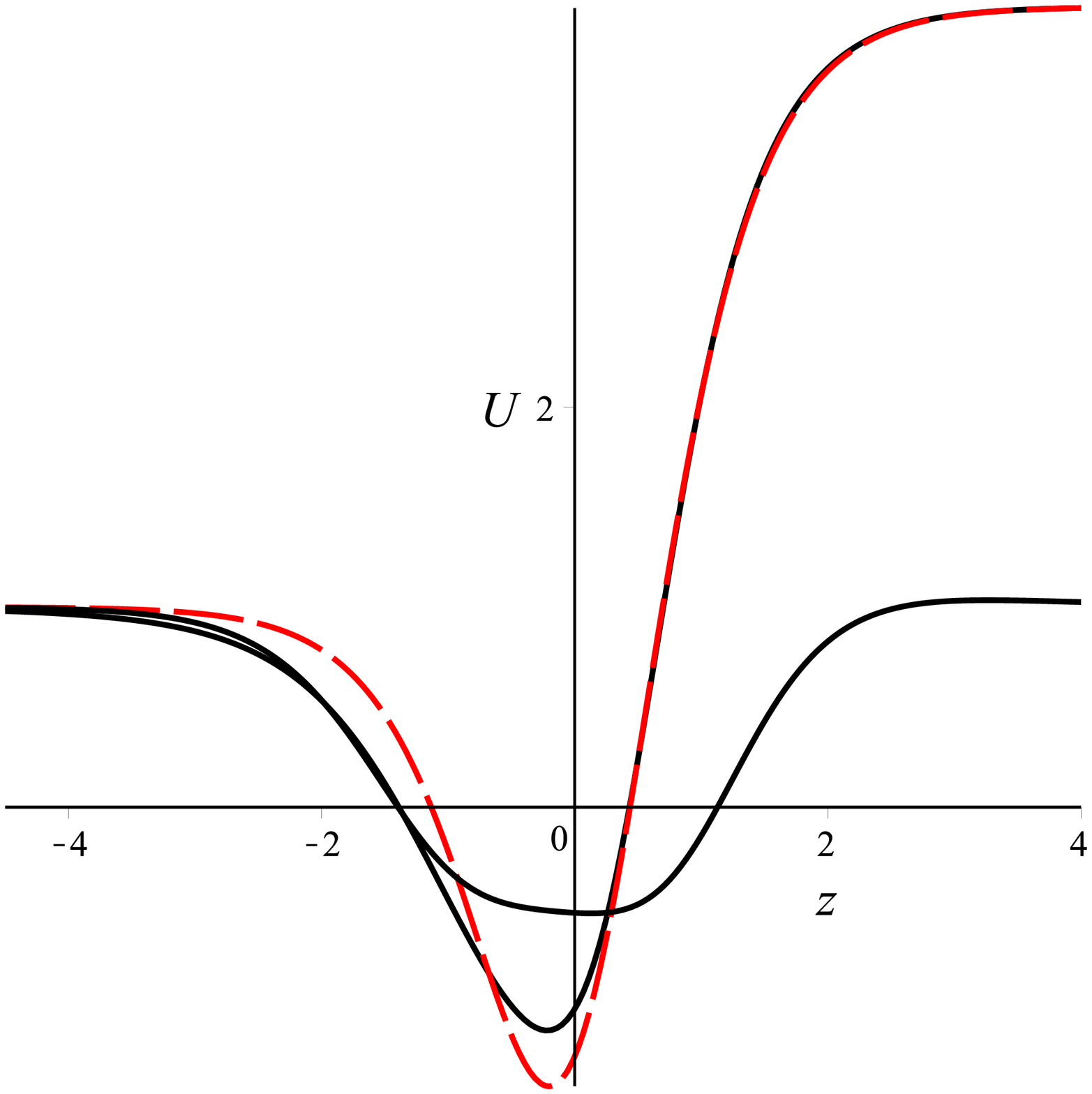}
\caption{The potential \eqref{V3}, the kink solutions \eqref{p3}, the energy density $\rho$, and the stability potential $U(z)$. For the same set of parameters chosen in Fig.~\ref{plot2}.}
\label{plot3}
\end{figure}

\begin{figure}%
\centering
\includegraphics[scale=0.3]{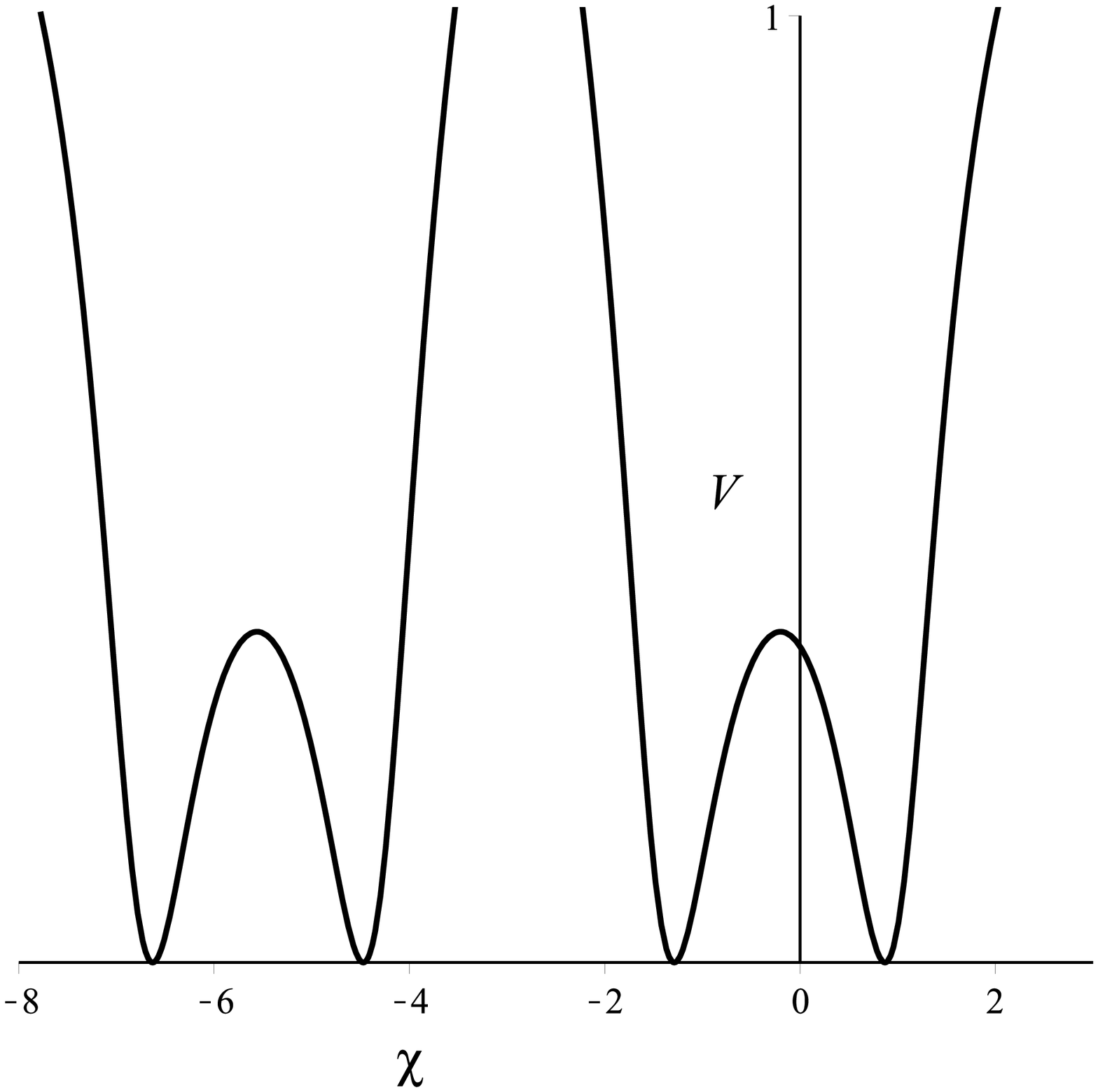}
\includegraphics[scale=0.3]{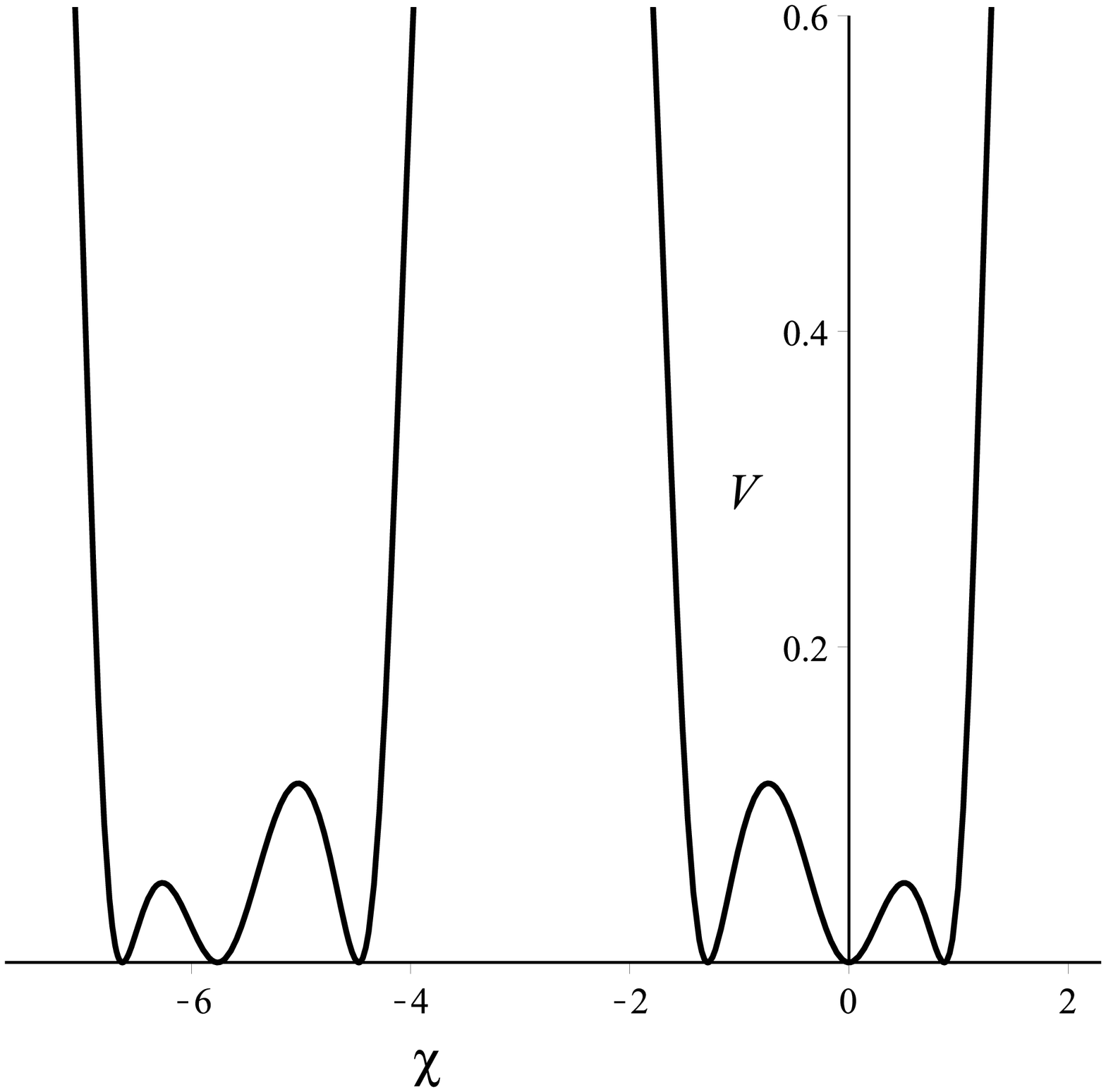}
\caption{The novel DBI models \eqref{V2} and \eqref{V3}, pictured in the left and right panels respectively, 
 using ${\alpha} = 1.2$.}
\label{plot4}
\end{figure}

The second deformation function is $f_2=2f_1^2-1$, that is $f_2(\chi)=2\chi^2\left(1+\chi/4\alpha^2 \right)^2-1$. The function $S_2(\chi)$ is
\be
S_2(\chi)=\chi\left(1+\frac{\chi}{4\alpha^2} \right) \left[\frac{1-\chi^2\left(1+{\chi}/{4\alpha^2} \right)^2}{1+{\chi}/{2\alpha^2}}\right].
\ee
It can also be written in terms of the first deforming function, that is $S_2(\chi)=f_1(\chi)S_1(\chi)$. And the deformed potential will be
\be
\label{V3}
V(\chi)= \alpha^2\left(1-\frac{1}{\sqrt{1+f_1(\chi)^2S_1(\chi)^2/\alpha^2}}\right).
\ee
The minima are obtained through $f_1(\chi_{min})S_1(\chi_{min})=0$, which are $\chi_{min}=0,-4\alpha^2,2\alpha^2(-1\pm\sqrt{1\pm1/\alpha^2})$. In the case $\alpha=1$, there are five minima localized at the points $0,-2,-4,$ and $-2(1\pm\sqrt{2})$; and so four topological sectors. Then the potential behaves like a standard $\phi^{10}$ model, as shown in the first graph of Fig.~\ref{plot3}. 

As the parameter $\alpha$ increases, one part of the  potential tends to approximate to the $\phi^6$ model, in such a way that for $|\alpha|>>1$, the standard $\phi^6$ model is restored. At the same time, another part of the potential is displaced along the negative $\chi$-axis. Such behaviour is presented in Fig.~\ref{plot4}, including the model previous seen.

The solutions are found by $\chi(x)=f_2^{-1}(\tanh(x))$, which are
\be
\label{p3}
\chi(x)=2\alpha^2\left(-1\pm \sqrt{1\pm\frac{1}{\alpha^2}\sqrt{\frac{1\pm\tanh(x)}{2}}}\right).
\ee
Real valued solutions require $|\alpha|\geq 1$. In the limit  $|\alpha|>>1$, the $\phi^6$ solutions are recovered which are $\chi_{6}(x)=\pm\sqrt{\frac{1\pm\tanh(x)}{2}}$.

The energy densities are calculated through \eqref{rhofo}, where
\be
\chi'(x)=\frac{\sech^2(x)}{4 \chi_{6}(x)\sqrt{1+\chi_{6}(x)/\alpha^2}},
\ee
and $\chi_{6}(x)$ is the static solution of the standard $\chi^6-$model.  

Figure~\ref{plot3} shows the model \eqref{V3} and its kink solutions (leftmost panels) for some values of $\alpha$. Also, it shows the energy densities and stability potentials (rightmost panels) for the larger and smaller sectors of the potential. In this situation, the DBI modifications do not induce a plateau on the stability potential, as happened on the previous models. The effect here is a flattening on the stability potential of the larger sectors.

\begin{figure}%
\centering
\includegraphics[scale=0.4]{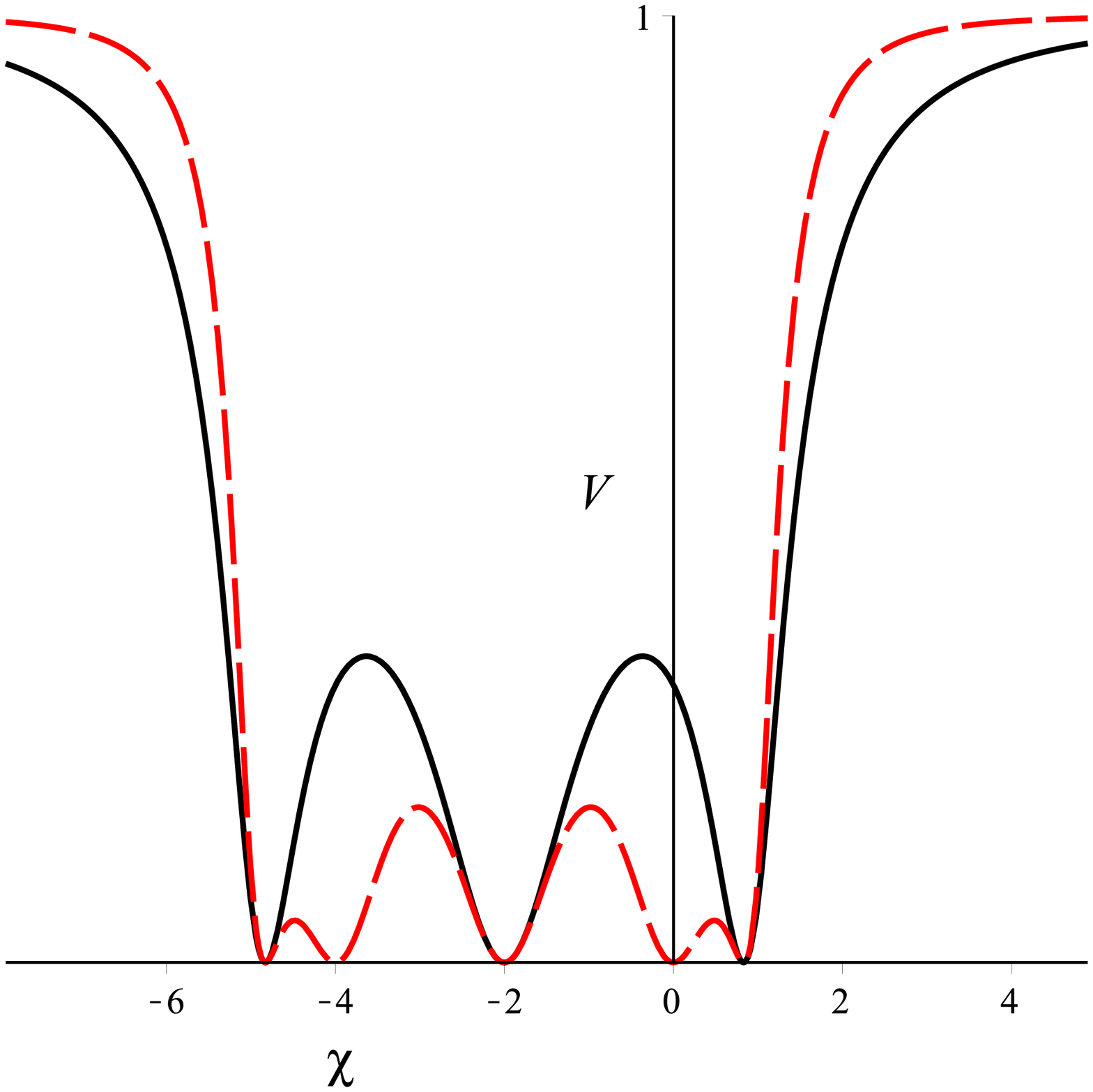}
\includegraphics[scale=0.4]{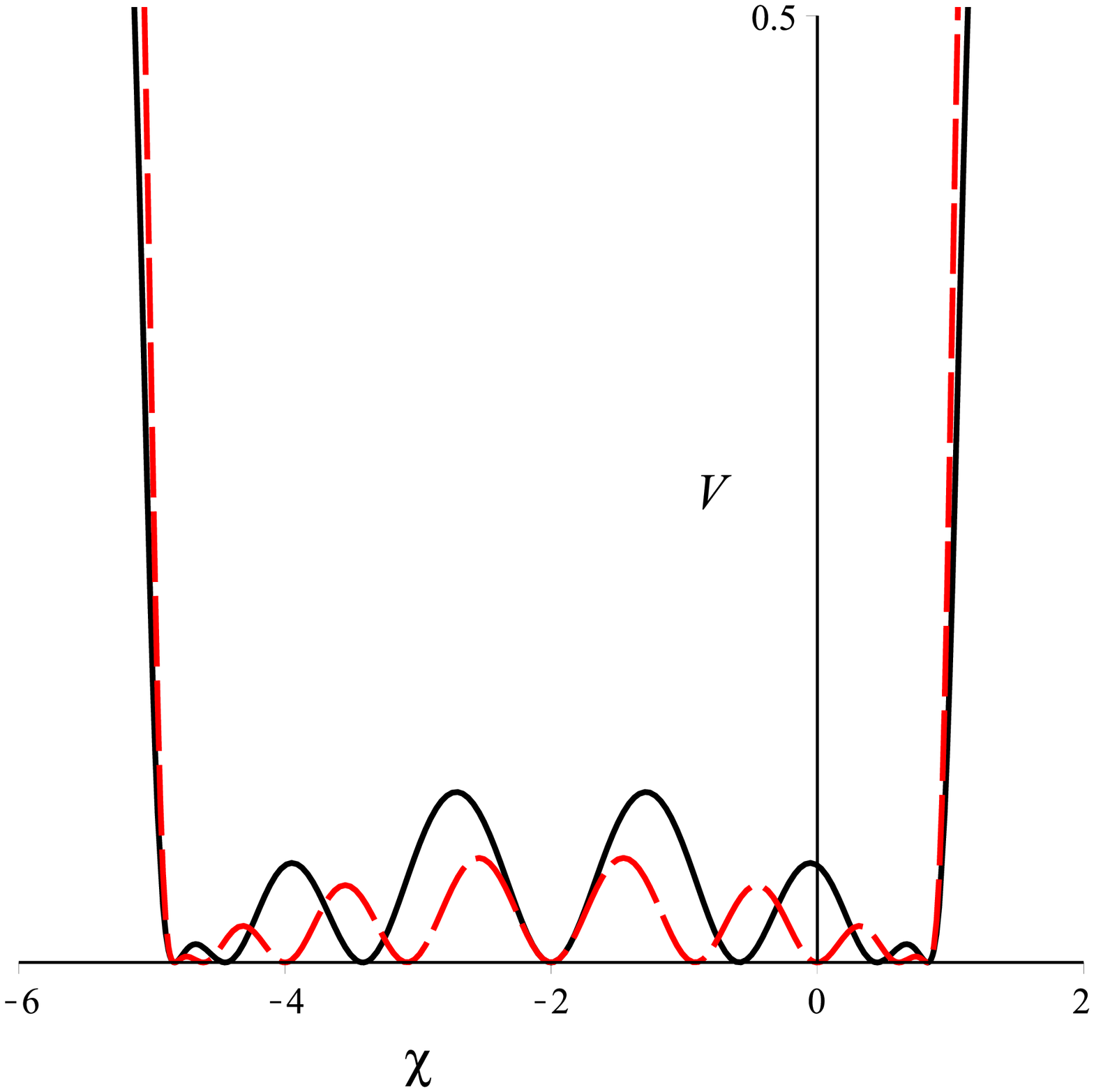}
\caption{Family of DBI potentials \eqref{Vn} fixing $\alpha=1$. In the left panel $n=1,2$; and in the right panel $n=3,4$.}
\label{plot5}
\end{figure}

These results can be extended for a family of potentials, where the relatives of standard models are approached by a DBI-like perspective. Let us select the general deforming function $f_n=\cos(n\arccos(f_1)-m\pi)$, with $n$ non-null positive integer, $m$ integer, and  $f_1(\chi)=\chi\left(1+\chi/4\alpha^2 \right)$ \cite{bazeiaLeon}.  The modified $S_{n}(\chi)$ functions are such that 
\be
S_{modified}(\chi)=\frac{S_{standard}(\chi\rightarrow f_1(\chi))}{f_{1,\chi}},
\ee
where $S_{standard}(\chi)$ are the $\alpha-$independent functions obtained at the scenarios \cite{bazeiaLeon,Bazeia:2017mnc}, and $f_{1,\chi}$ is the derivative of $f_1(\chi)$ with respect to its argument. Therefore
\be
S_{n}(f_n)=\frac{(1-f_1^2)}{n f_{1,\chi}}U_{n-1}(f_1),
\ee 
that is
\be
S_{n}(\chi)=\frac{1}{n}S_1(\chi)U_{n-1}\left(\chi\left(1+\chi/4\alpha^2 \right)\right),
\ee
where $S_1(\chi)$ is given by expression~\eqref{S1}, and $U_{n-1}$ are the Chebyshev polynomials of second kind. Particularly,   $n=1,2$ constitute the last modified polynomial models presented in this section. 

The deformed potentials are given by
\be
\label{Vn}
V_n(\chi)= \alpha^2\left(1-\frac{1}{\sqrt{1+\frac{1}{\alpha^2 n^2} S_1^2(\chi)U_{n-1}^2\left(\chi\left(1+\chi/4\alpha^2 \right)\right)}}\right),
\ee
which is represented in Fig.~\ref{plot5} for some values of $n$, and $\alpha=1$. The minima are
\be
\chi_{n,l}^{min}=-2\alpha^2\left(1\pm\sqrt{1+\frac{1}{\alpha^2}\cos\left(\frac{l\pi}{n}\right)}\right)
\ee
with $l=0,1,..,n$. The solutions are
\be
\chi_{n,m}(x)=2\alpha^2\left(-1\pm \sqrt{1+\frac{1}{\alpha^2}\cos\left(\frac{\theta(x)+m\pi}{n}\right)}\right),
\ee
where $\theta(x)=\arccos(\tanh(x))$ and $m=0,..,2n-1$. The energy density is obtained by 
\be
\chi'_{n,m}(x)=\frac{\sin\left(\frac{\theta+m\pi}{n}\right)\sech(x)}{n\sqrt{1+\frac{1}{\alpha^2}\cos\left(\frac{\theta+m\pi}{n}\right)}},
\ee
and the stability potential reads 
\ben
U(x)=\frac{1}{S_1U_{n-1}}\frac{d}{dx}\left( 
\frac{S_1U_{n-1}\left\{\left(S_{1,\chi}+(1-n)f_1\right) U_{n-1}+nU_{n-2} \right\}}{n\left(1+\frac{1}{\alpha^2n^2}S_1U_{n-1}^2\right)^2}  \right). \nonumber \\
\een

\begin{figure}%
\centering
\includegraphics[scale=0.3]{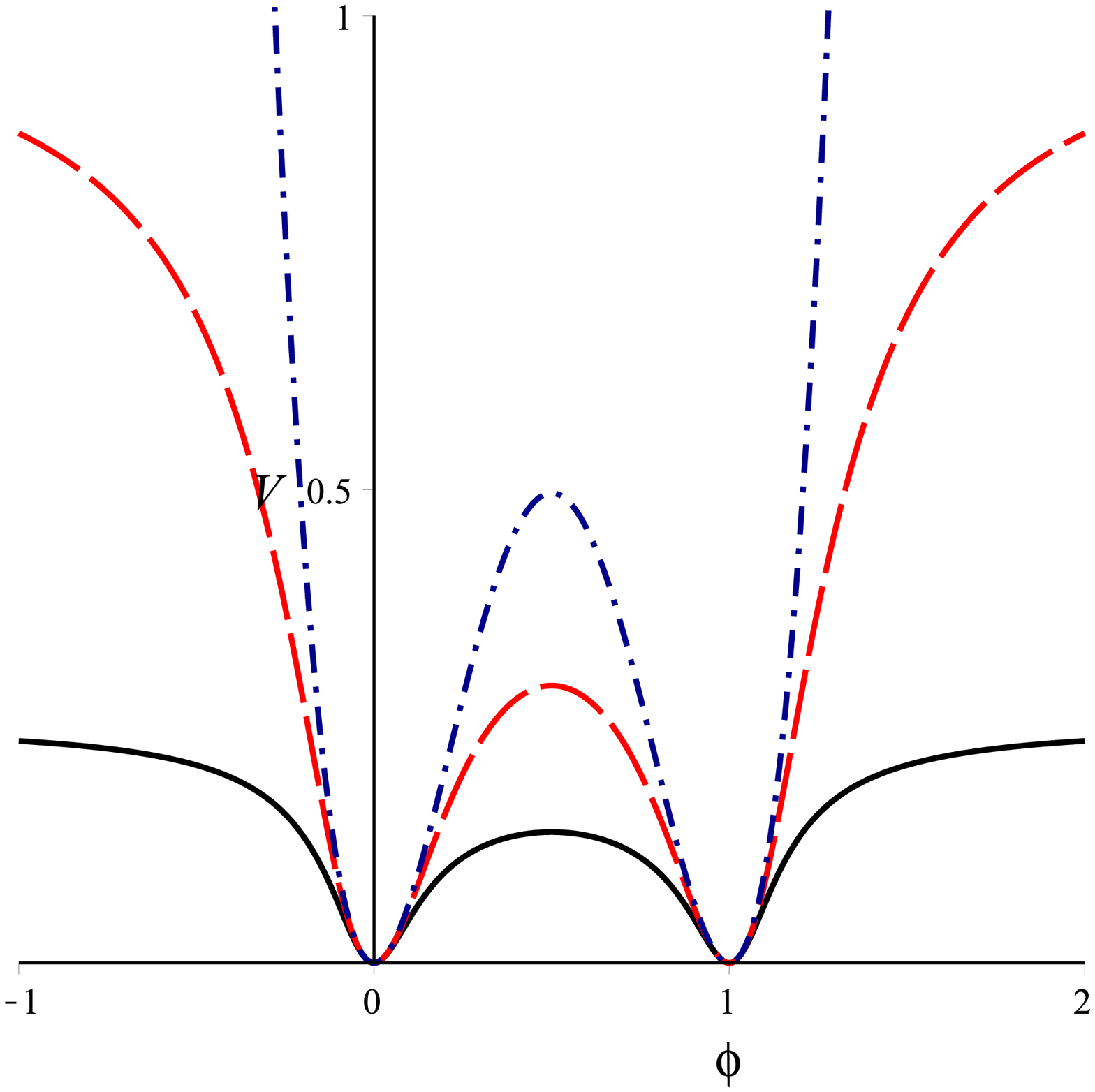}
\includegraphics[scale=0.3]{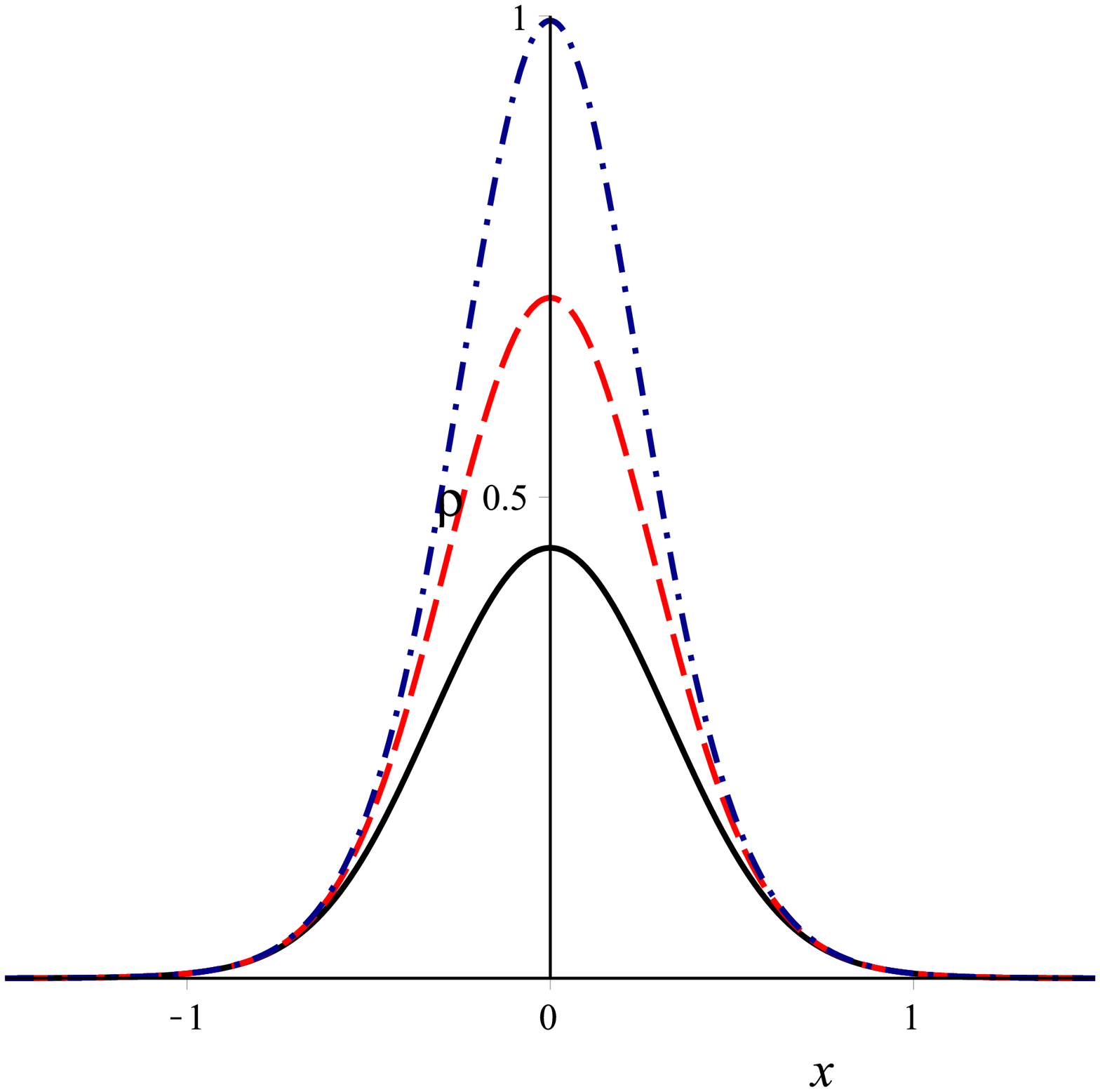}
\includegraphics[scale=0.3]{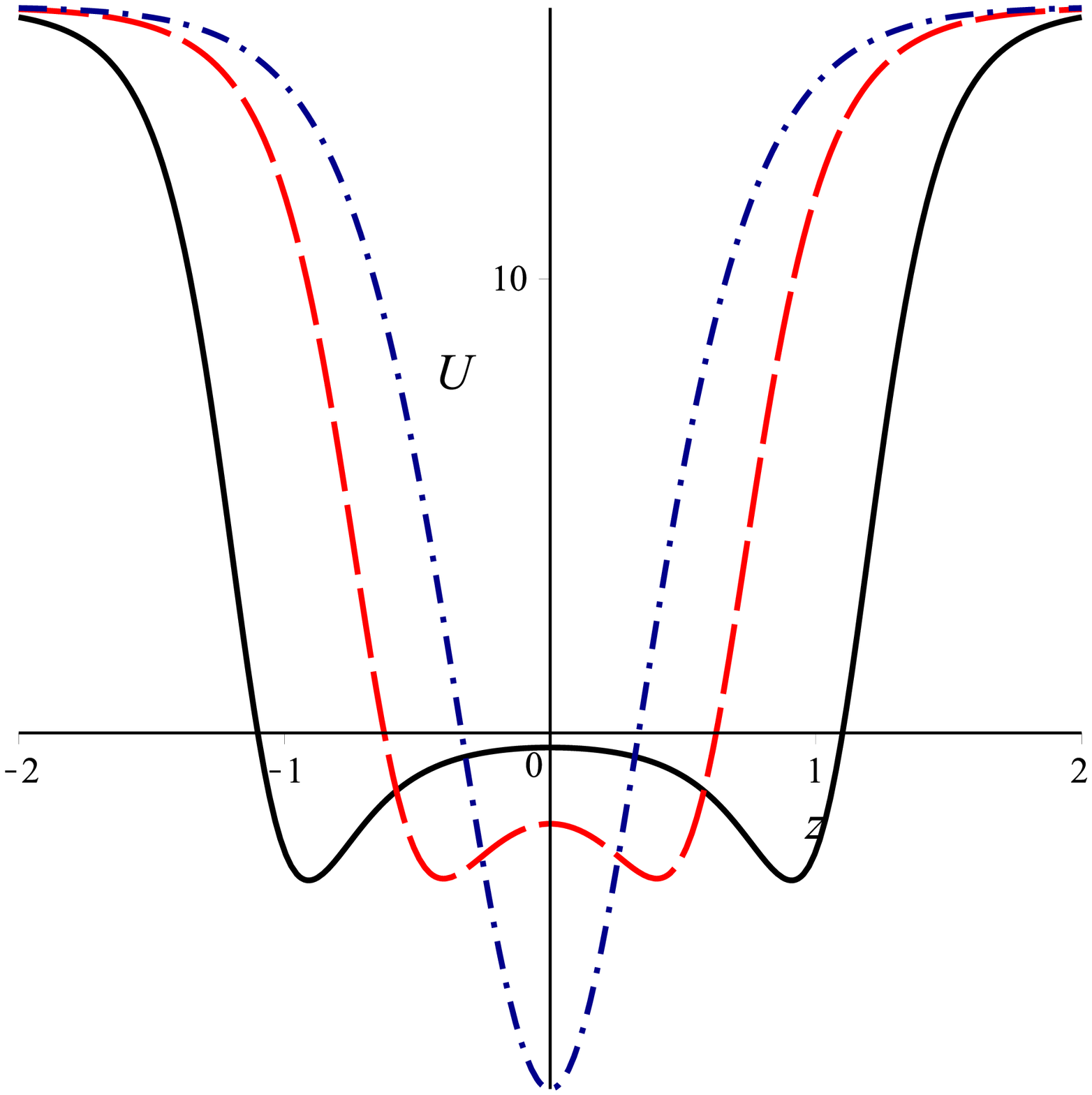}
\caption{The potential \eqref{pot1staro} (left panel), the energy densities (centre panel) and stability potentials (right panel) for the same values of $\alpha$ used in Fig.~\ref{plot1}.}
\label{fig3a}
\end{figure}

\subsection{Models relatives of the $n-$Starobinsky}

\subsubsection{The first generalized $n-$Starobinky}

\begin{figure}%
\centering
\includegraphics[scale=0.3]{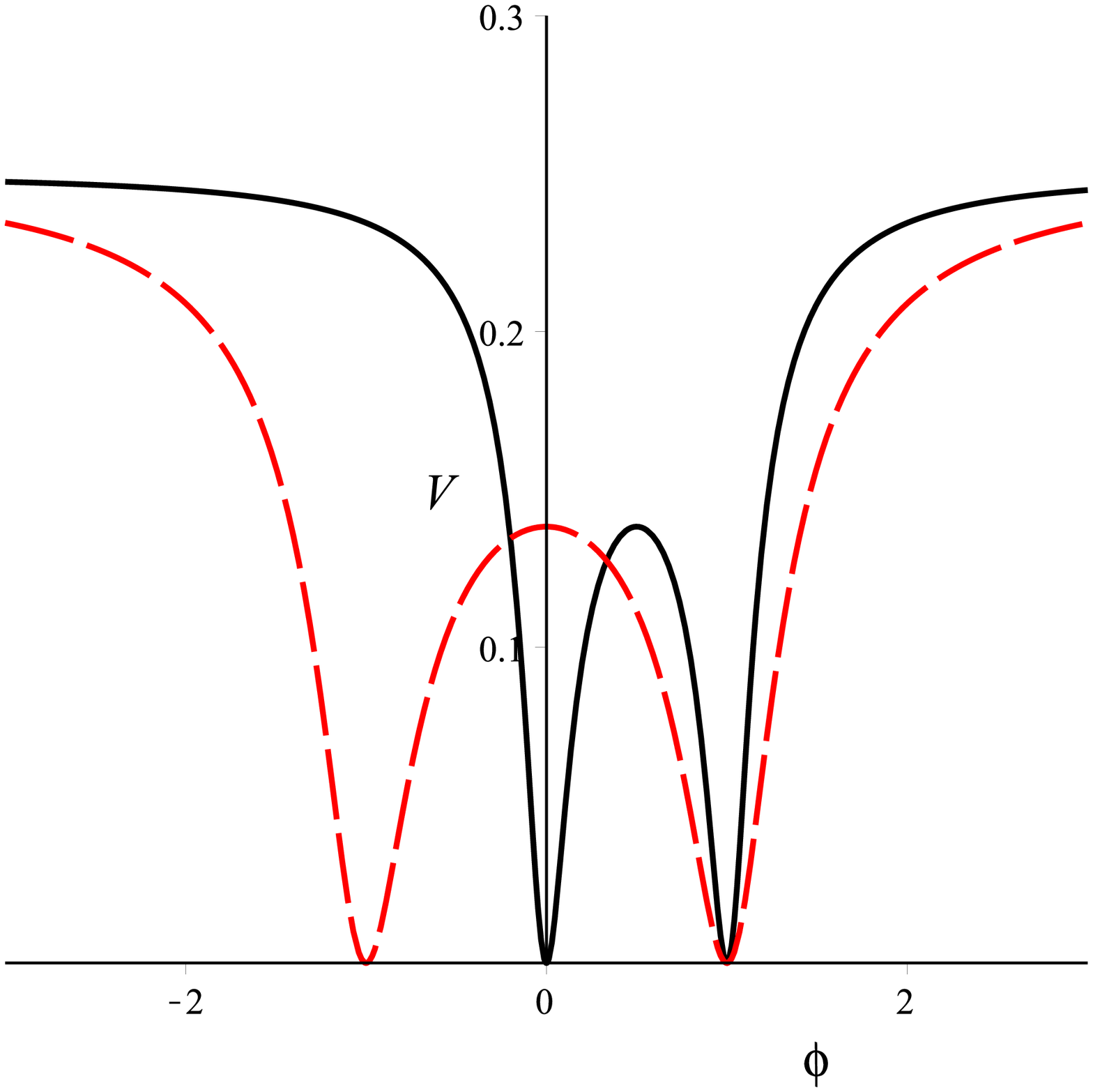}
\includegraphics[scale=0.3]{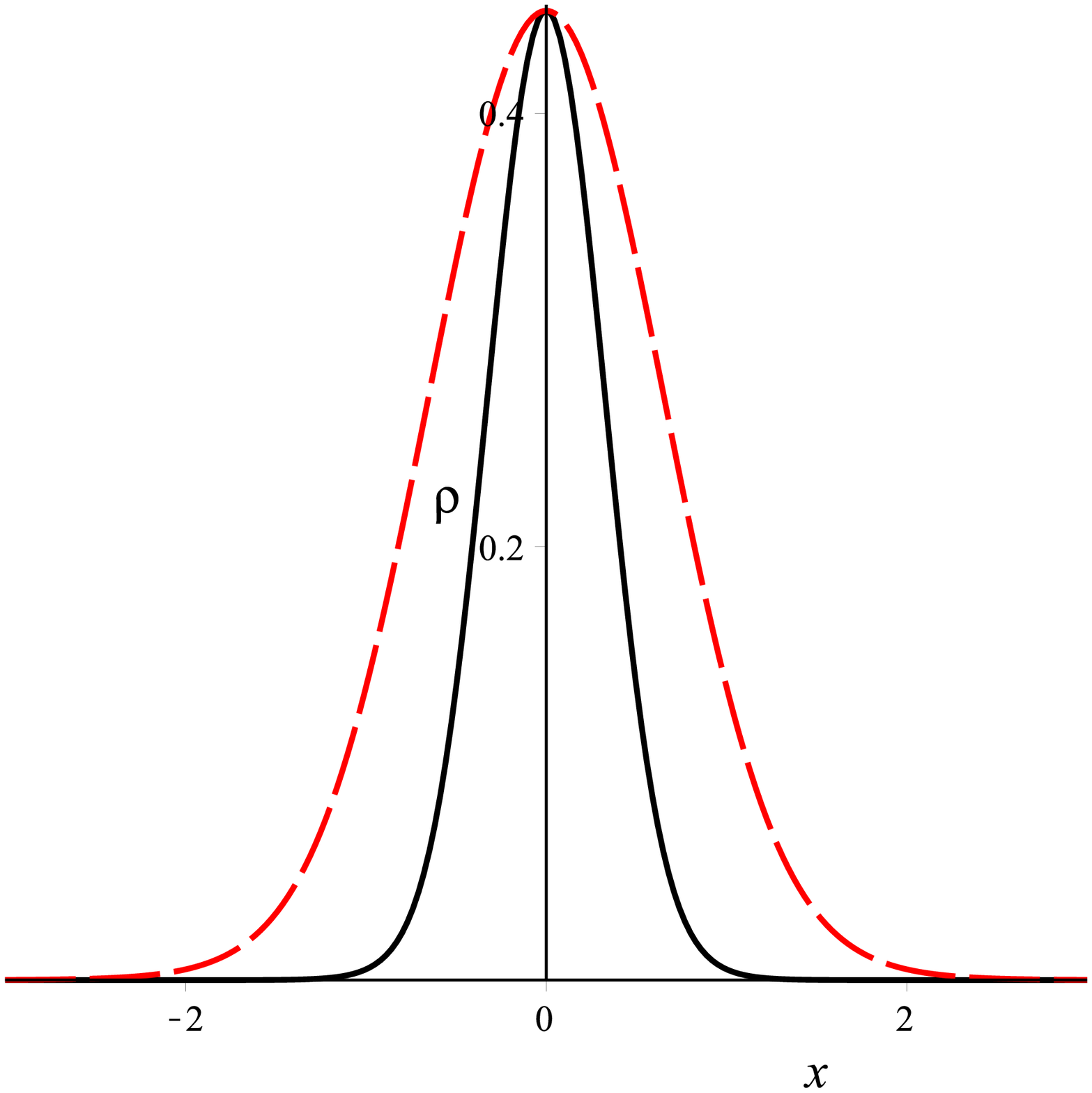}
\includegraphics[scale=0.3]{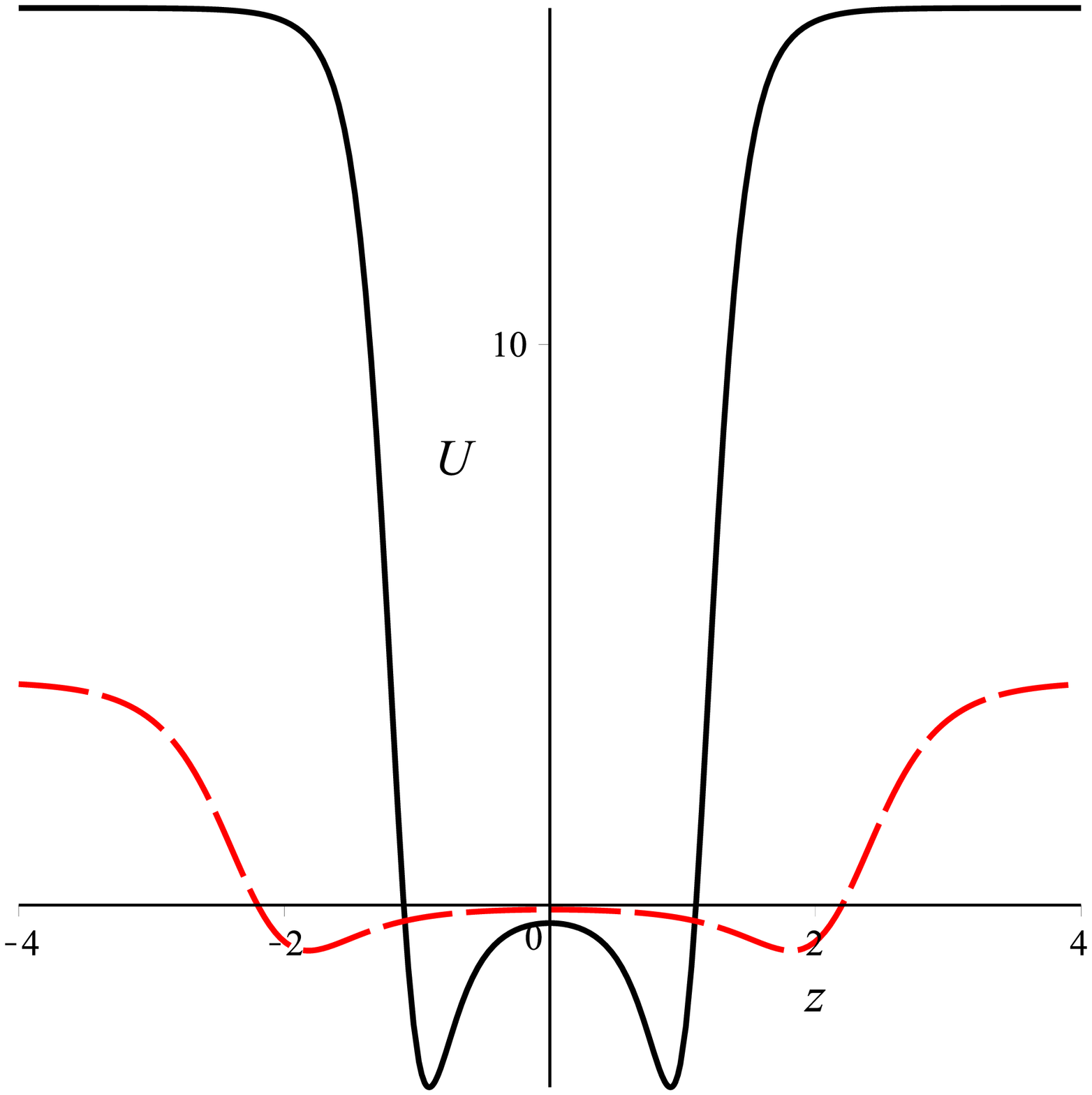}
\caption{A comparative between the modified models \eqref{pot1} and \eqref{pot1staro} fixing $\alpha=0.5$, represented respectively by dashed (red) and solid (black) lines. }
\label{fig3b}
\end{figure}

In this subsection, we are going to propose a first modification on the standard $n-$Starobinsky potential. Initially, we suggest
\be
\label{pot1staro}
V(\phi)=\alpha^2\left(1-\frac{1}{\sqrt{1+\dfrac{16}{\alpha^2}\phi^2(1-\phi)^2}}\right),
\ee
which will be our deforming model used to get a more general $n-$Starobinsky. In this case, $\phi'=4\phi(1-\phi)$ and the solutions are the same obtained by the standard scenario, Eq.~\eqref{sol1staro}. The energy density is 
\be
\label{rh1star}
\rho(x)=\frac{\sech^4(2x)}{\sqrt{1+\frac{1}{\alpha^2}\sech^4(2x)}},
\ee
and the stability potential is obtained by
\be
F_{X}V_\phi=-\frac{4\,\sech^2(2x)\tanh(2x)}{\left(1+\frac{1}{\alpha^2}\sech^2(2x)\right)^2}.
\ee
Fig.~\ref{fig3a} shows how this model behaves as $\alpha$ varies. In Fig.~\ref{fig3b}, the same quantities are compared with the ones found by the $\phi^4$ relative \eqref{pot1} for a small $\alpha$. The kink solutions are already compared on the second panel in Fig.~\ref{fig2}.  If  $\alpha$ is large, the results become the same seen in Fig.~\ref{fig2} for the standard situation.

\begin{figure}%
\centering
\includegraphics[scale=0.3]{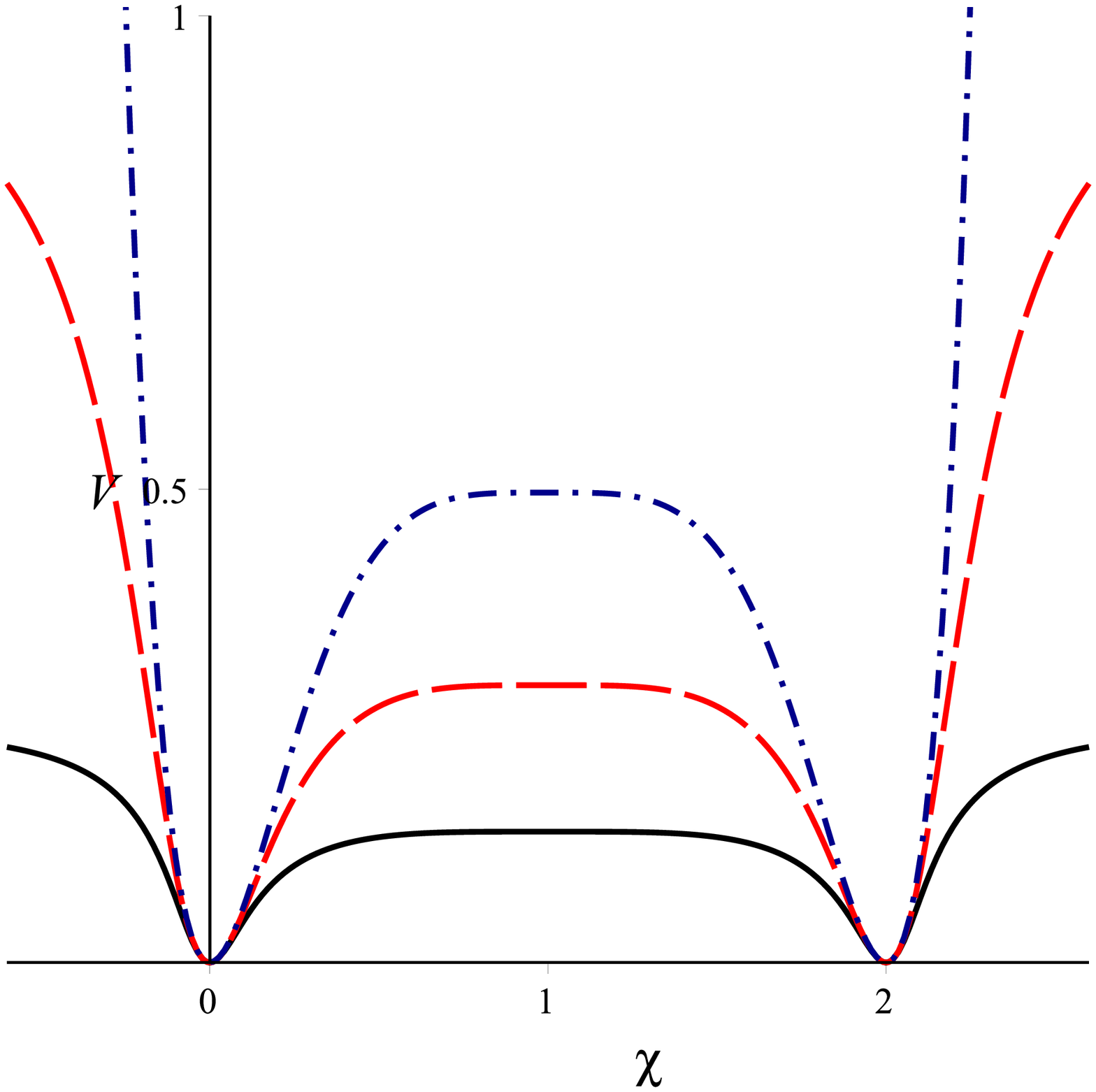}
\includegraphics[scale=0.3]{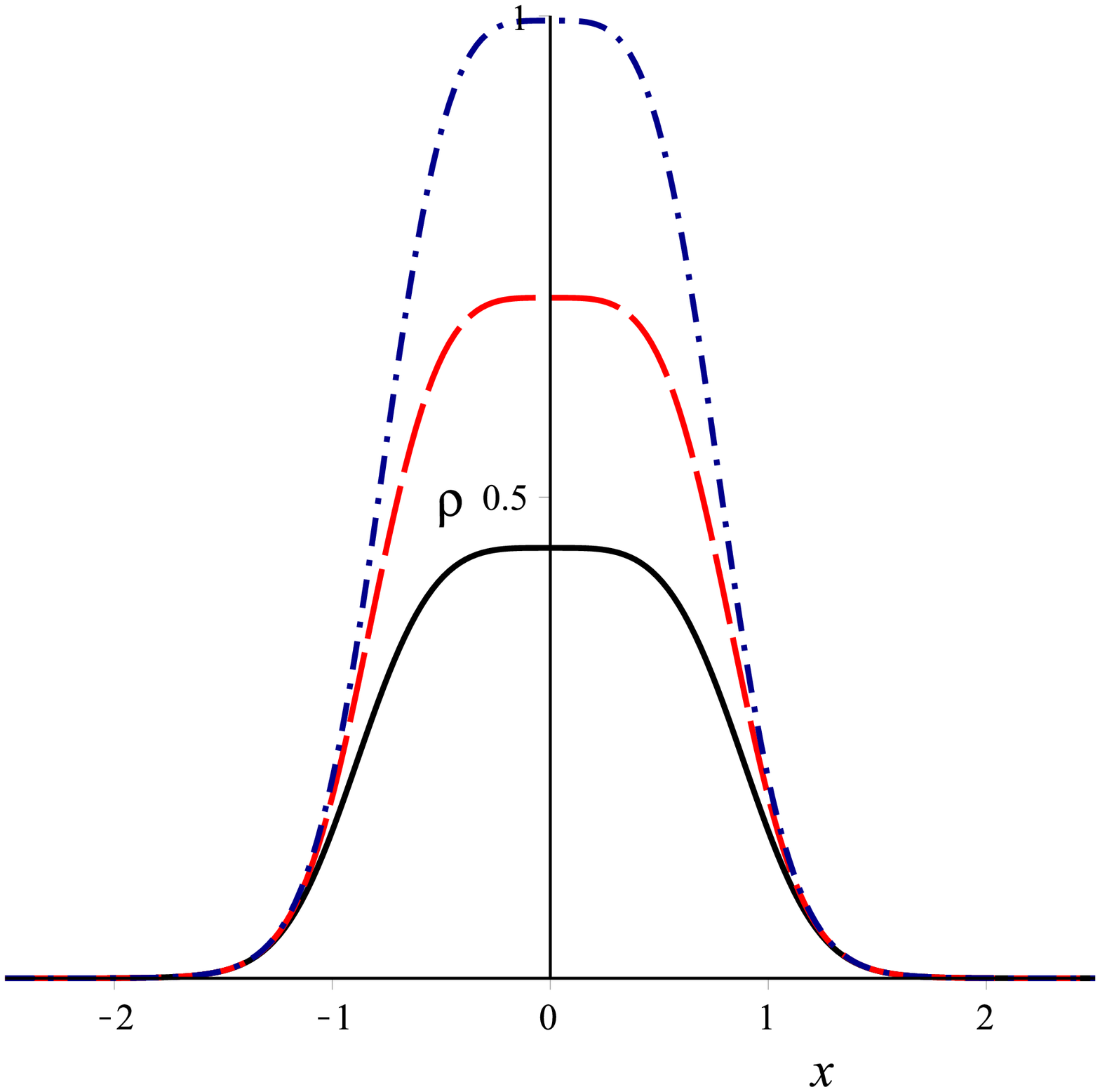}
\includegraphics[scale=0.3]{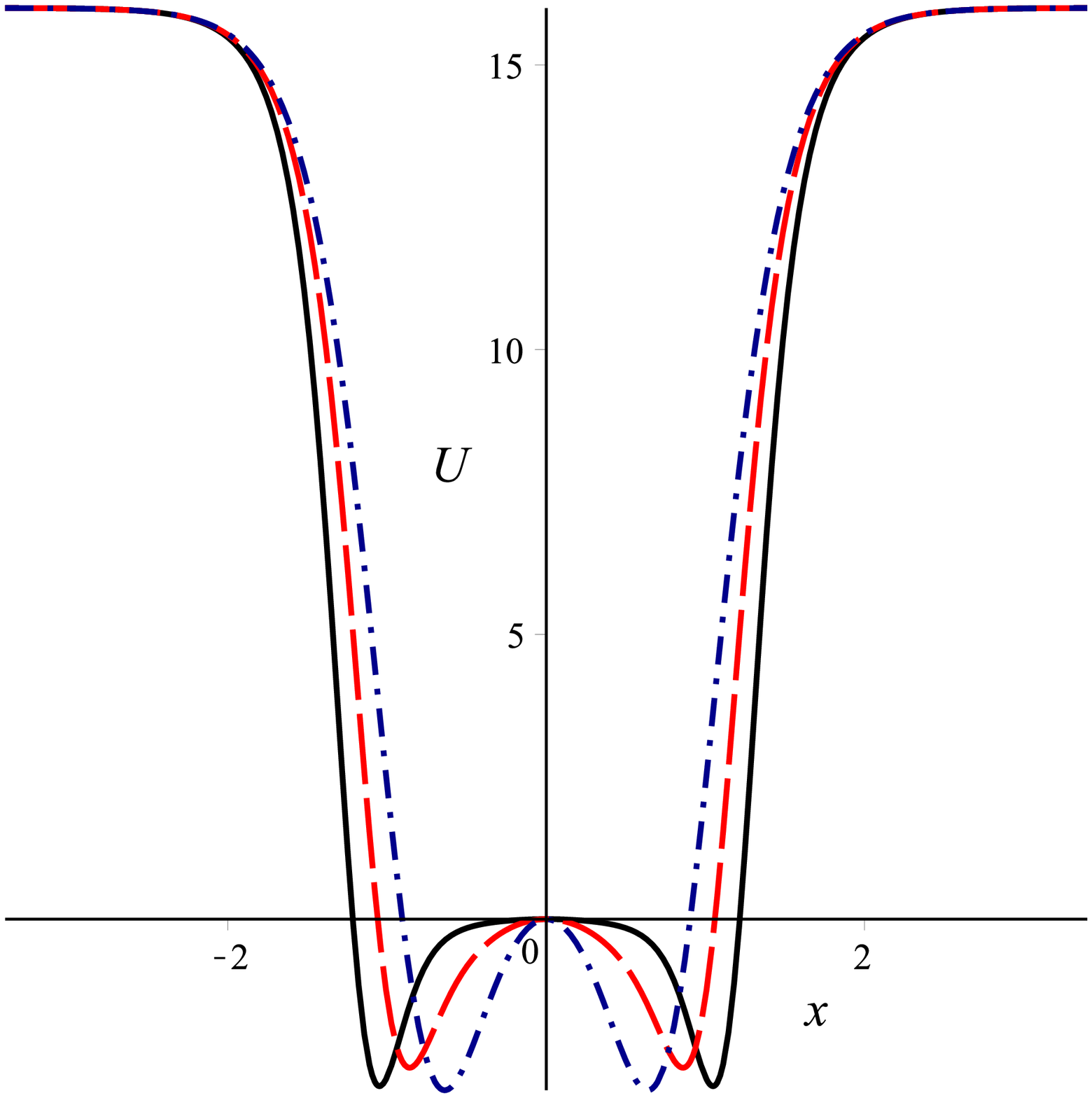}
\caption{The deformed $n-$Starobinsky potential \eqref{pot2staro}, varying $\alpha$ and fixing $n=2$, for the same $\alpha$ values used in Fig.~\ref{plot1}.}
\label{fig4}
\end{figure}

From that, it is possible to propose a set of modified $n-$Starobinsky models, by deforming the previous model 
\be
S(\chi)=\frac{4f(1-f)}{f_\chi}.
\ee
Through the same function adopted in the standard case  \eqref{funcstaro}, we have
\be
\label{S1star}
S(\chi)=1-\left(1-2\chi/n\right)^{2n},
\ee
and the deformed potential \eqref{DP} becomes
\be
\label{pot2staro}
V_n(\chi)=\alpha^2\left(1-\frac{1}{\sqrt{1+\frac{1}{\alpha^2}\left(1-\left(1-\frac{2}{n}\chi\right)^{2n}\right)^2}}\right),
\ee
which has the same minima and maxima found by the ordinary $n-$Starobinsky. The topological solutions also are the same found in the standard scenario, Eqs.~\eqref{hyptranscen} and \eqref{lerchitranscen}. The energy density and stability potential are modified according to the DBI dynamics. The case $n=1$ is shown in Fig.~\ref{fig3a}, and the case $n=2$ is shown in Fig.~\ref{fig4}, for some $\alpha$ values.

\begin{figure}%
\centering
\includegraphics[scale=0.4]{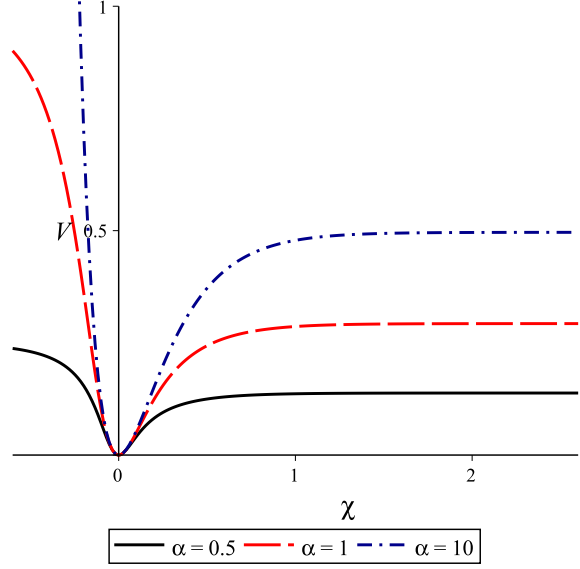}
\caption{The first $\alpha-$Starobinsky  potential $V_\alpha(\chi)$, Eq.~\eqref{alphastar}, varying the parameter $\alpha$.}
\label{fig5}
\end{figure}

The deformed potential \eqref{pot2staro} constitutes a generalization of the $n-$Starobinsky model, as can be seen through the expansion $\alpha>>1$
\ben
V_n(\chi)= \frac12\left(1-\left(1-\frac{2}{n}\chi\right)^{2n}\right)^2 -\frac{3}{8\alpha^2}\left(1-\left(1-\frac{2}{n}\chi\right)^{2n}\right)^4+ {\cal O}\left(\frac{1}{\alpha^4}\right). \nonumber \\
\een
At the limit $n\rightarrow \infty$, the expression \eqref{pot2staro} provides the following modification on the original Starobinsky model
\be\label{alphastar}
V_\alpha(\chi)=\alpha^2\left(1-\frac{1}{\sqrt{1+\frac{1}{\alpha^2}\left(1-\e^{-4\chi}\right)^2}}\right);
\ee
for $\alpha>>1$, the original Starobinsky model is recovered at leading order
\be
V(\chi)=\frac{1}{2}\left(1-\e^{-4\chi}\right)^2-\frac{3}{8\alpha^2}\left(1-\e^{-4\chi}\right)^4+{\cal O}\left(\frac{1}{\alpha^4}\right),
\ee
as suggested on Fig.~\ref{fig5}.

\subsubsection{The second generalized $n-$Starobinsky}

\begin{figure}%
\centering
\includegraphics[scale=0.22]{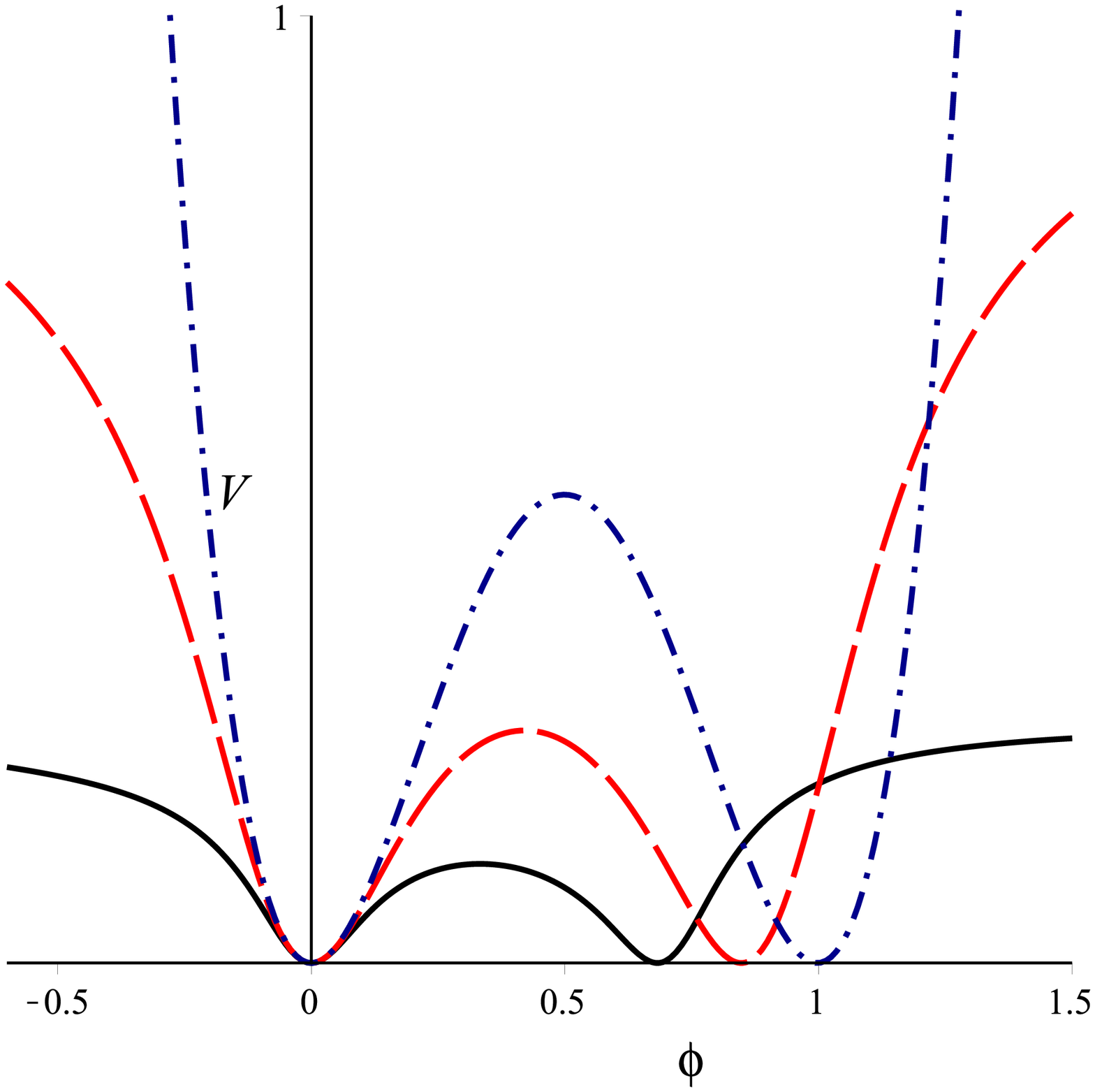}
\includegraphics[scale=0.22]{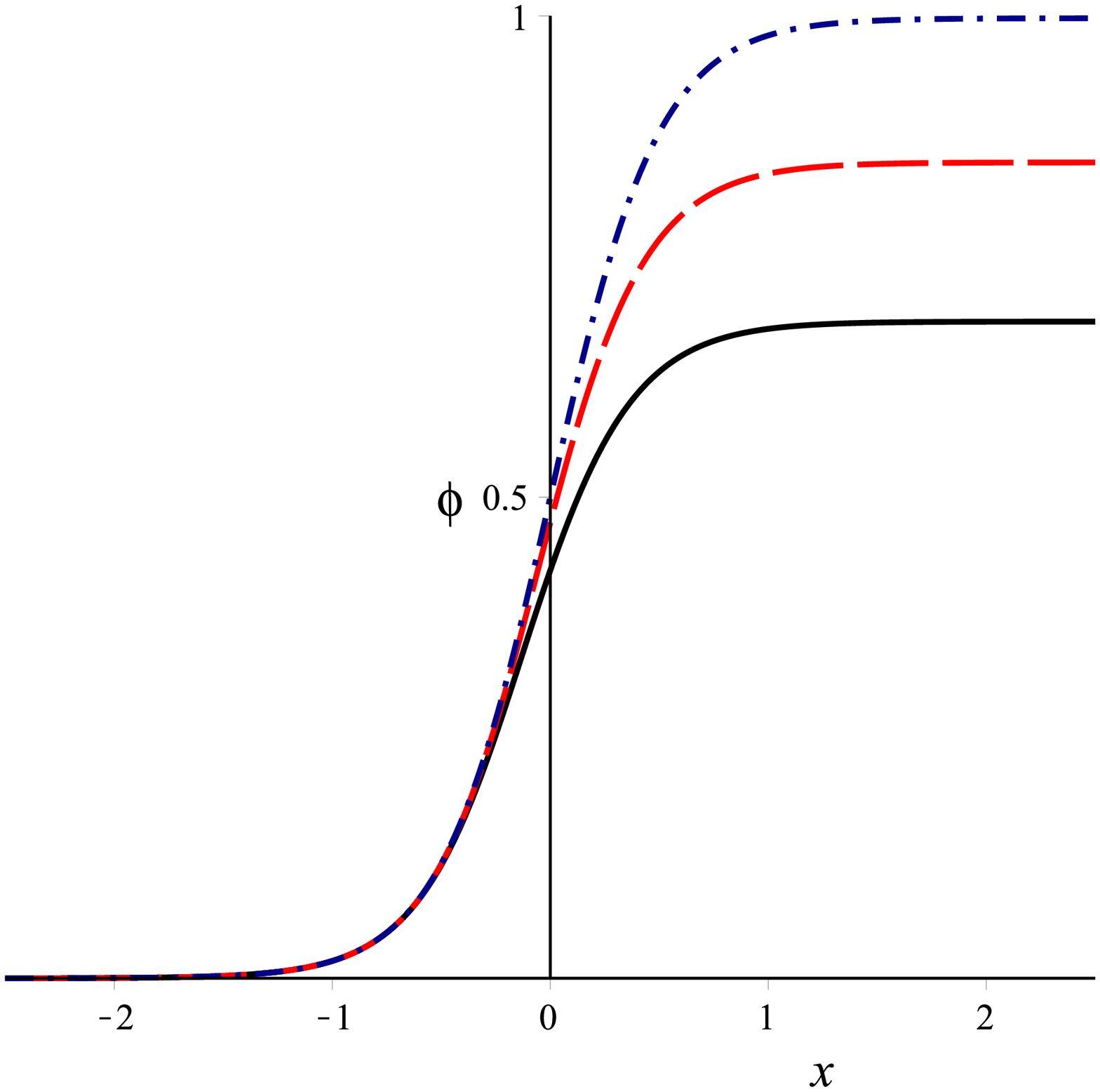}
\includegraphics[scale=0.22]{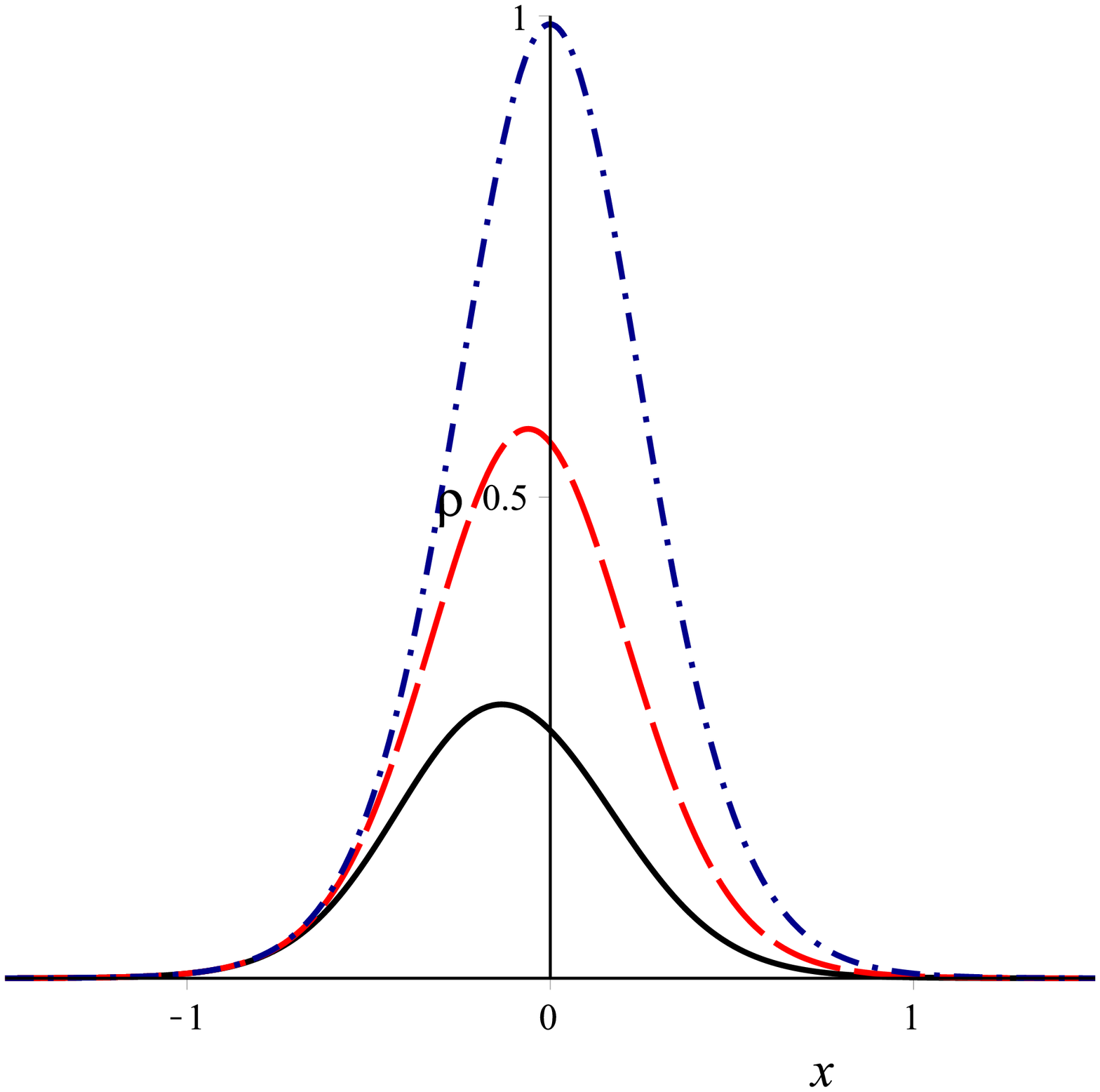}
\includegraphics[scale=0.22]{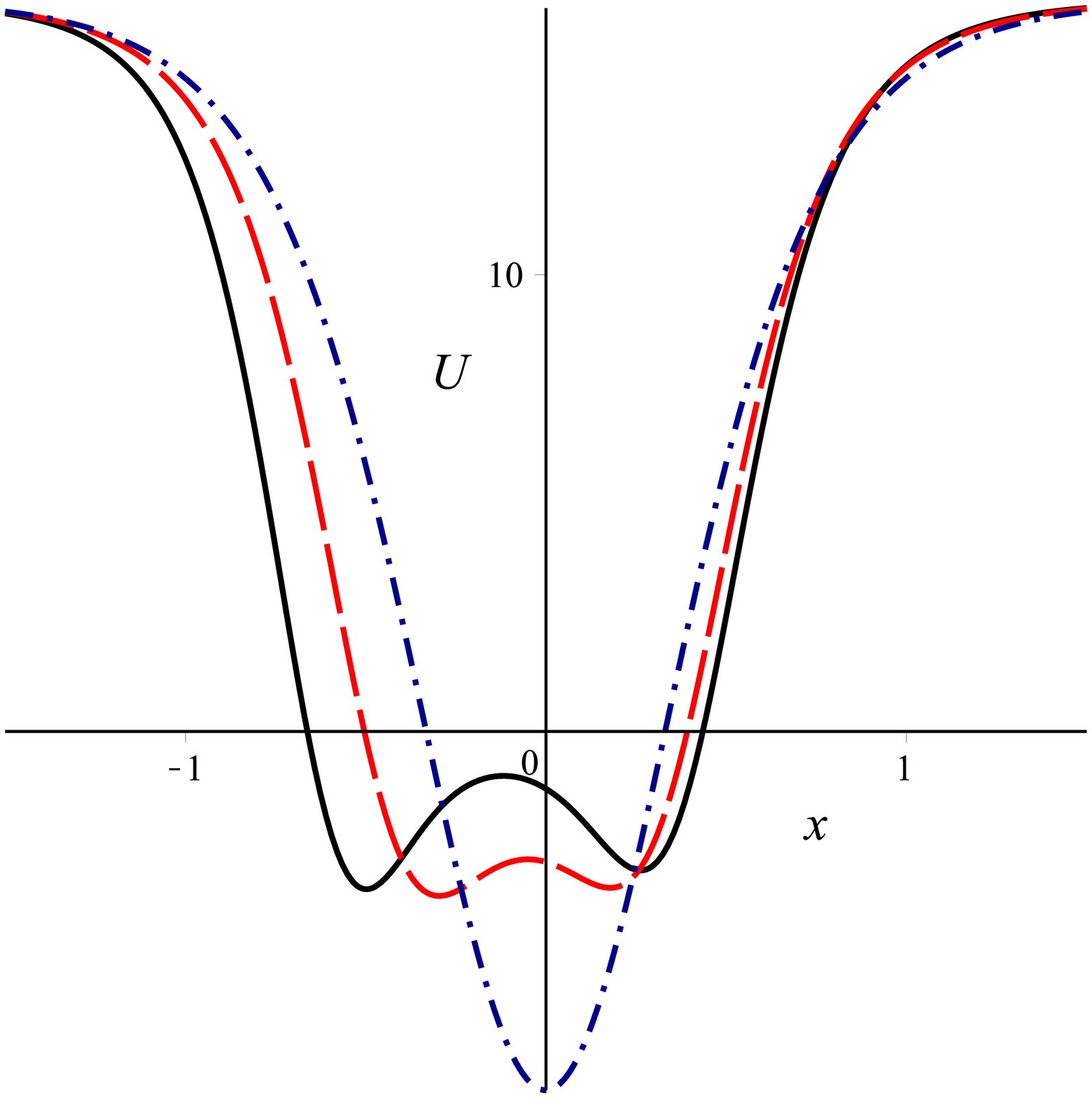}
\caption{The second generalized $n-$Starobinsky model for $n=1$, and the same $\alpha$ values used in Fig.~\ref{plot1}. The potentials $V_{n,\alpha}(\chi)$, Eq.~\eqref{secondstaro} (first panel),  their kink solutions \eqref{lerchitranscen2} (second panel), the energy densities (third panel), and the stability potentials (fourth panel).}
\label{fig6}
\end{figure}

\begin{figure}%
\centering
\includegraphics[scale=0.22]{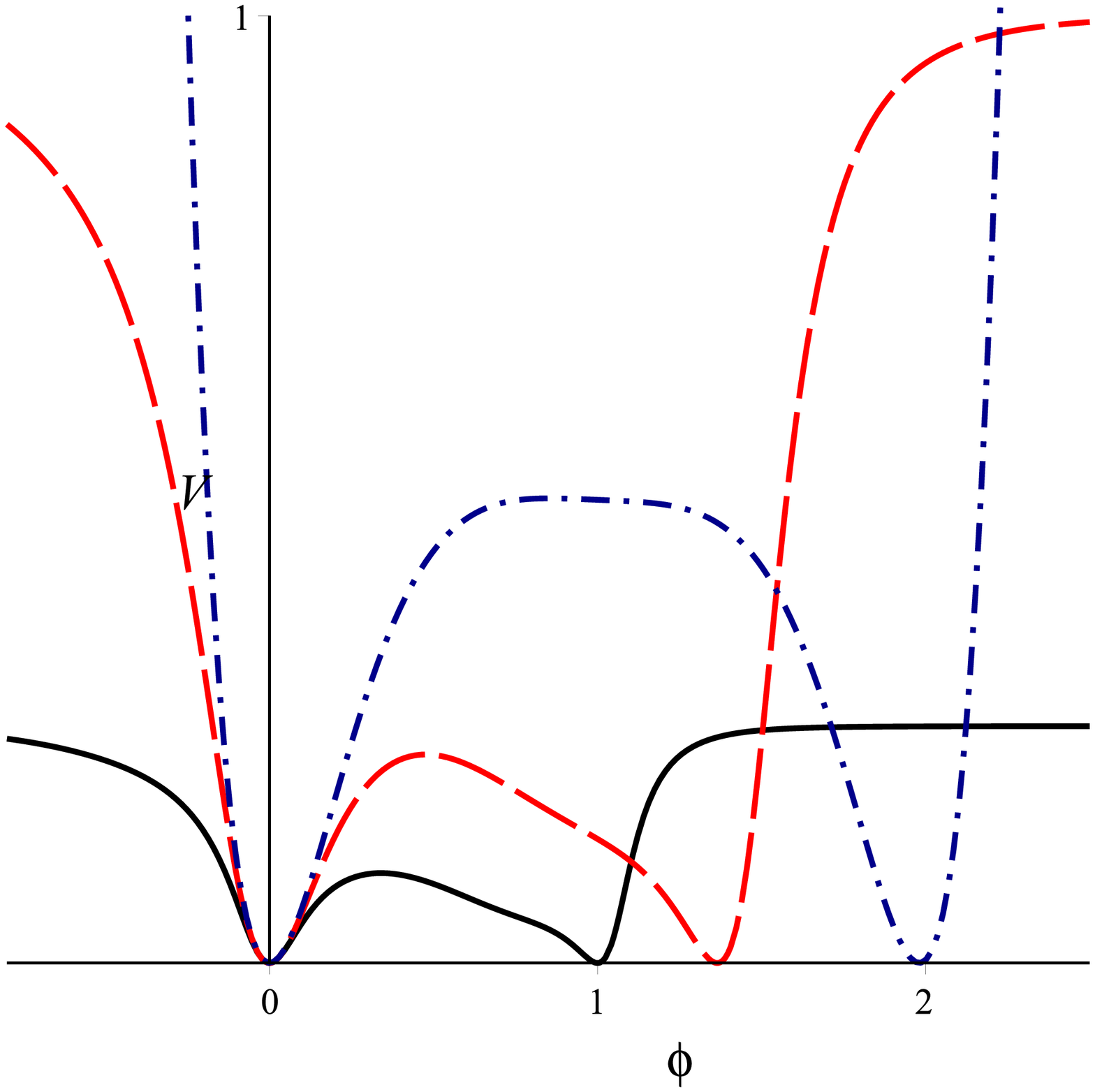}
\includegraphics[scale=0.22]{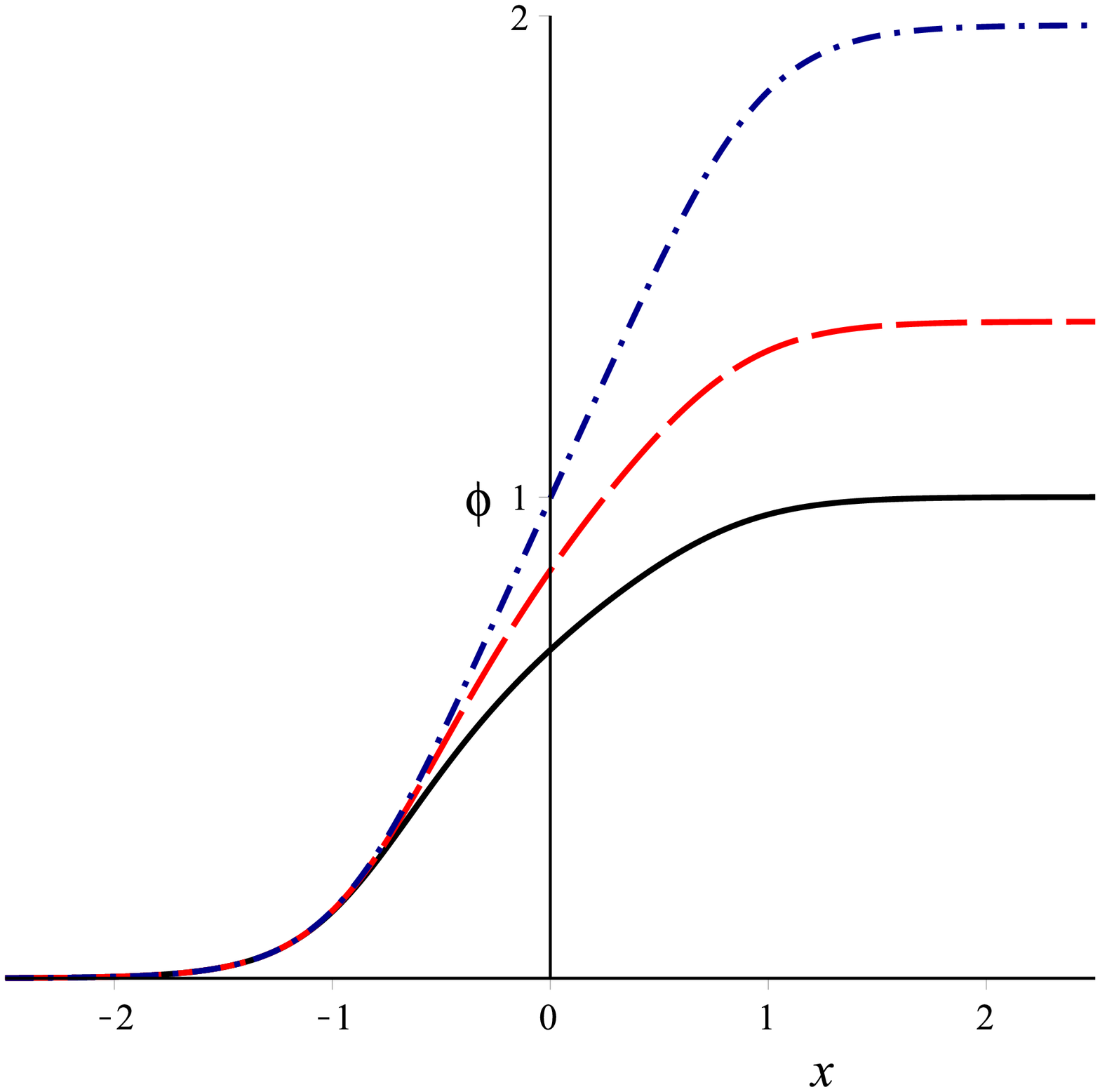}
\includegraphics[scale=0.22]{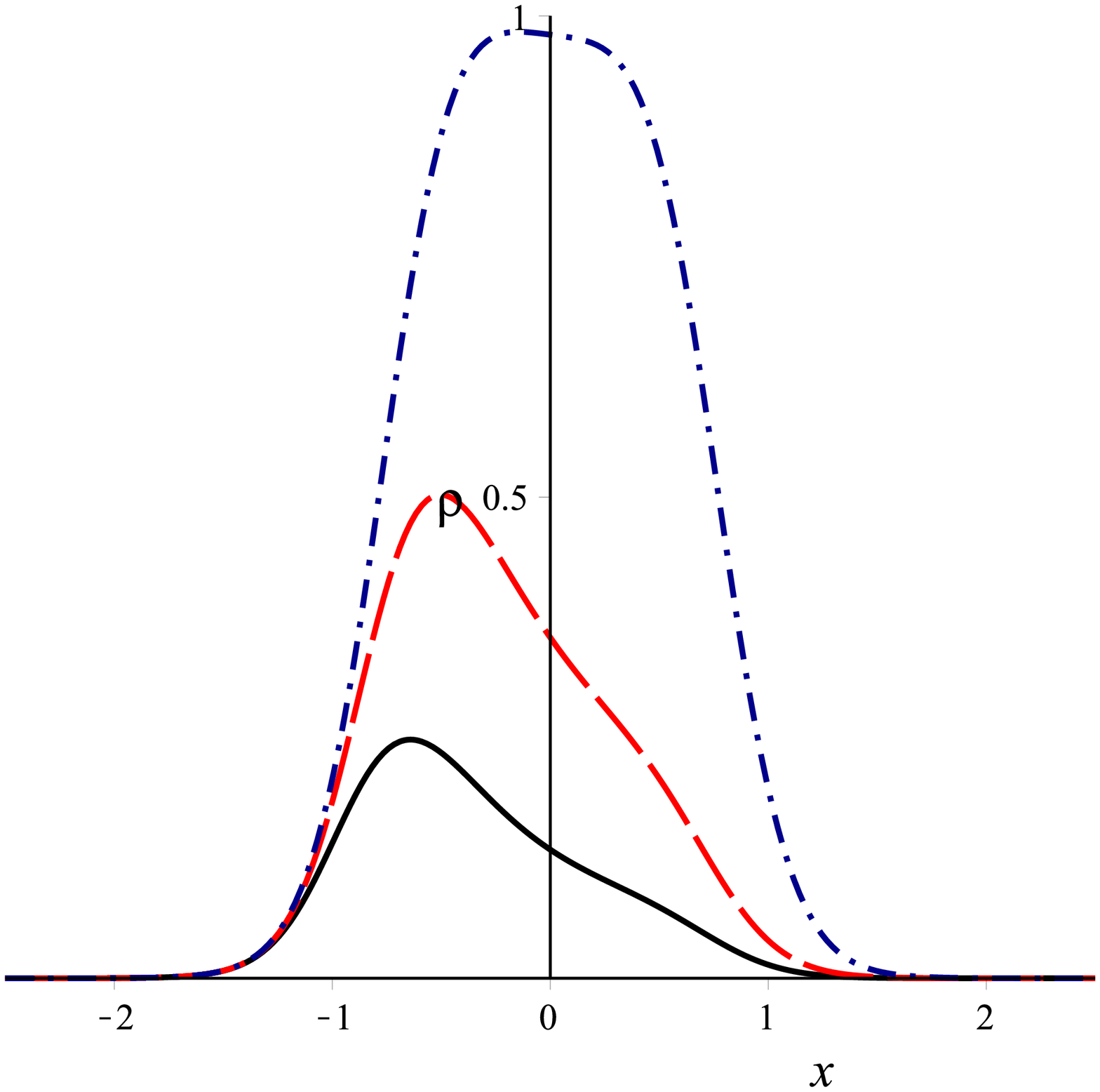}
\includegraphics[scale=0.22]{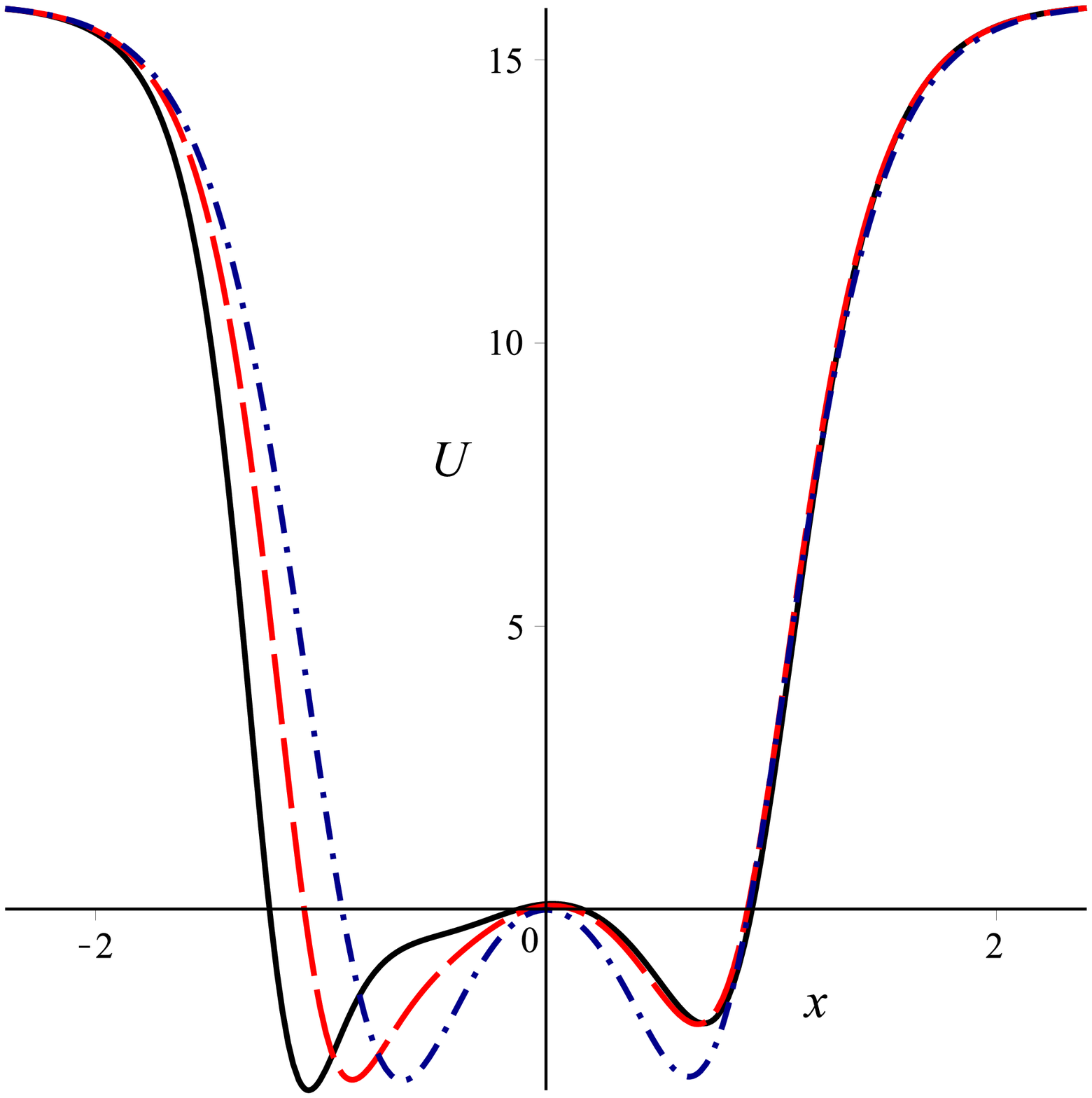}
\caption{The second generalized $n-$Starobinsky model for $n=2$, and the same quantities adopted in Fig.~\ref{fig6}.}
\label{fig7}
\end{figure}

Another possible generalization of the $n-$Starobinsky model is made applying a deformation $\chi \rightarrow f(\phi)$ on the  previous general model \eqref{pot2staro}, then
\be
S_n(\phi)=\frac{1-\left(1-\frac{2}{n}f\right)^{2n}}{f_\phi}.
\ee
Selecting $f(\phi) = \phi(1+\phi^2/4\alpha^2)$, we have
\be
S_n(\phi)=\frac{1-\left(1-\frac{2}{n} \phi\left(1+\frac{\phi^2}{4\alpha^2}\right) \right)^{2n}}{1+3\phi^2/4\alpha^2};
\ee
and the novel modified $n-$Starobinsky potential is
\ben
\label{secondstaro}
V_{n,\alpha}(\phi) = \alpha^2\left(1-\frac{1+{3\phi^2}/{4\alpha^2}}{\sqrt{\left(1+\frac{3\phi^2}{4\alpha^2}\right)^2+\frac1{\alpha^2}
\left(1-\left(1-\frac{2}{n} \phi\left(1+\frac{\phi^2}{4\alpha^2}\right) \right)^{2n}\right)^2}}\right). \nonumber \\
\een
It forms another generalization of the standard $n-$Starobinsky that is restored when $\alpha\rightarrow\infty$.

The kink solutions can be obtained by the substitution of $\chi \rightarrow f(\phi)$ in the transcendental solution \eqref{lerchitranscen}, that is 
\ben
\label{lerchitranscen2}
\left(1-\frac2n\phi\left(1+\frac{\phi^2}{4\alpha^2}\right)\right)\Phi\left(\left(1-\frac2n\phi\left(1+\frac{\phi^2}{4\alpha^2}\right)\right)^{2n},1,\frac1{2n}\right)=-4x. \nonumber \\
\een

\begin{figure}%
\centering
\includegraphics[scale=0.4]{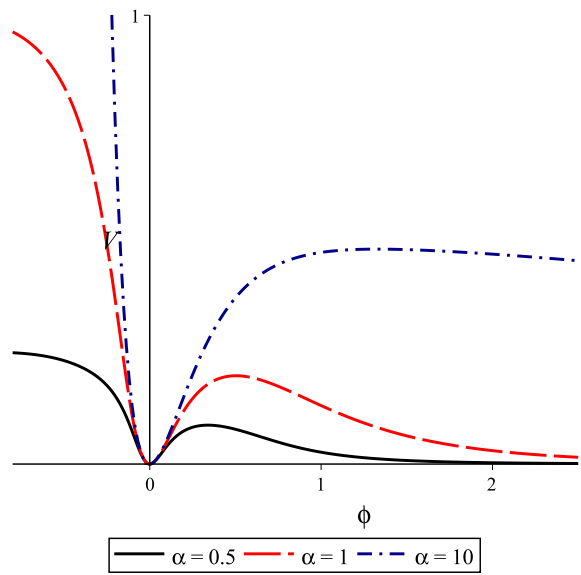}
\caption{The second $\alpha-$Starobinsky model, Eq.~\eqref{secalphastaro}, for the same $\alpha$ values adopted in Fig.~\ref{fig6}.}
\label{fig8}
\end{figure}

Figures~\ref{fig6} and \ref{fig7} show the behavior of the potentials $V_{n,\alpha}(\chi)$, the kink solutions, their energy densities and stability potentials for some values of $\alpha$, and $n=1,2$. As the parameter $\alpha$ increases, the results approach to the canonical $n-$Starobinsky; and for small $\alpha$, an asymmetrical behavior is observed.

On the other hand, the limit $n\rightarrow \infty$ furnishes the following modification on the original Starobinsky model
\ben
\label{secalphastaro}
V_{\alpha}(\phi) = \alpha^2\left(1-\frac{1+{3\phi^2}/{4\alpha^2}}{\sqrt{\left(1+\frac{3\phi^2}{4\alpha^2}\right)^2+\frac1{\alpha^2}
\left(1-\e^{-4\phi\left(1+\frac{\phi^2}{4\alpha^2}\right)}\right)^2}}\right), \nonumber \\
\een
which we are calling of second $\alpha-$Starobinsky model. As suggested in Fig.~\ref{fig8}, the ordinary Starobinsky is  precisely recovered for $\alpha>>1$.

\section{Comments and conclusions}\label{sec-com}

In this paper we have addressed the existence of scalar field solutions in DBI dynamics in the presence of several potentials and their relatives, i.e., their deformed counterparts. 
The mechanism may find several applications from topological solutions that can be considered as kinks in two-dimensional spacetimes to domain walls or braneworlds in higher dimensions, such as four or five-dimensional spacetimes \cite{brane}. Furthermore, up to properly changing in the coordinates, the static one-dimensional spatial solutions can be analytical continued to provide cosmological solutions --- see e.g., \cite{cvetic-soleng} in four and \cite{bbc} in five-dimensions. Thus the scalar potentials of the original theory have inflaton potentials counterparts in cosmology and their form may satisfy the requirements to describe observational cosmology, mainly in the inflationary phase. Interestingly enough we have found some relatives of the Starobinsky model that approaches the shape found in the KKLT potential \cite{KKLT} which has been found through flux compactification in string theory. Several interesting investigations such as tunneling effects among the vacua can be considered in the present scenario to address the issue of vacuum stability. This can be crucial for both cosmology and bubble nucleation driven by 
potentials whose relatives have a barrier shape in the presence of a softly breaking term in the Lagrangian to render non-topological solutions \cite{kolb}. These are interesting points that merit further consideration in upcoming  studies.

\section*{Acknowledgments}

We would like to thank CNPq, CAPES and CNPq/PRONEX/FAPESQ-PB (Grant no. 165/2018), for partial financial support. FAB acknowledges support from CNPq (Grant no.  312104/2018-9). The author E.E.M Lima also thank to PRPGI/IFBA and PROEX/IFBA for their support.


\end{document}